\documentclass[
 reprint,
 amsmath,amssymb,
 aps,
]{revtex4-2}

\usepackage{graphicx}
\usepackage{dcolumn}
\usepackage{bm}
\usepackage{hyperref}

\usepackage[usenames,dvipsnames,svgnames,table]{xcolor}

\usepackage{xspace} 
\newcommand{\textcode}[1]{\textsc{#1}}
\newcommand{\kokkos}{\textcode{Kokkos}\xspace}
\newcommand{\Athenapp}{\textcode{Athena++}\xspace}
\newcommand{\Athena}{\textcode{Athena}\xspace}
\newcommand{\kathena}{\textcode{K-Athena}\xspace}

\newcommand{\Pm}{\mathrm{Pm}}

\def\Ms#1{{\texttt{Ms#1}}}

\def\Ms#1{{\texttt{Ms#1}}}
\def\Ma#1{{\texttt{Ma#1}}}
\def\MsMa#1#2{{\texttt{Ms#1\_Ma#2}}}

\begin{document}

\preprint{APS/123-QED}

\title{Magnetized Decaying Turbulence in the Weakly Compressible Taylor-Green Vortex}

\author{Forrest W. Glines}
\email[]{glinesfo@msu.edu}
\altaffiliation{Department of Computational Mathematics, Science, and Engineering, Michigan State University}
\affiliation{Department of Physics and Astronomy, Michigan State University}

\author{Philipp Grete}
\email[]{grete@pa.msu.edu}
\affiliation{Department of Physics and Astronomy, Michigan State University}

\author{Brian W. O'Shea}
\email[]{oshea@msu.edu}
\altaffiliation{Department of Computational Mathematics, Science, and Engineering and National Superconducting Cyclotron Laboratory, Michigan State University}
\affiliation{Department of Physics and Astronomy, Michigan State University}

\date{\today}

\begin{abstract}
Magnetohydrodynamic turbulence affects both terrestrial and astrophysical plasmas.
The properties of magnetized turbulence must be better understood to more accurately characterize these systems.
This work presents ideal MHD simulations of the compressible Taylor-Green vortex under a range of initial sub-sonic Mach numbers and magnetic field strengths.
We find that regardless of the initial field strength, the magnetic energy becomes dominant over the kinetic energy on all scales after at most several dynamical times.
The spectral indices of the kinetic and magnetic energy spectra become shallower than $k^{-5/3}$ over time and generally fluctuate.
Using a shell-to-shell energy transfer analysis framework, we find that the magnetic fields facilitate a significant amount of the energy flux and that the kinetic energy cascade is suppressed.
Moreover, we observe nonlocal energy transfer from the large scale kinetic energy to intermediate and small scale magnetic energy via magnetic tension.
We conclude that even in intermittently or singularly driven weakly magnetized systems, the dynamical effects of magnetic fields cannot be neglected.
\end{abstract}

\maketitle

\section{Introduction}

Magnetized turbulence is present in many terrestrial and astrophysical plasmas. 
Turbulence in magnetohydrodynamics (MHD) has been studied extensively over recent decades, from experimental, theoretical, and numerical perspectives, as the field continues to work towards a full understanding of magnetized turbulent plasmas.
However, much of the theoretical and numerical work focuses on continuously driven plasmas, where a continuous (although potentially stochastic) force adds energy to the plasma, resulting in stationary turbulence.
In many natural systems, the turbulence can be intermittently driven by infrequently occurring events or initialized from the initial conditions.
For example, in the circumgalactic medium (CGM), the hot diffuse gas surrounding galaxies, or in the intracluster medium (ICM), the plasma in galaxy cluster that accounts for the majority of baryonic mass, turbulence can be introduced by various mechanisms. These include mergers with other galaxies, brief increases in the birth rate of stars, temporary outflows from jets driven by gas accreting onto supermassive black holes, supernovae, and many more transient events \citep{normanClusterTurbulence1999,larsonTurbulenceStarFormation1981,britzenNewView872017,korpiSupernovaregulatedInterstellarMedium1999-fixed}.
In pulsed power plasmas such as in a z-pinch, the plasma is driven by a single initial event and then allowed to decay into turbulence as kinetic and magnetic energy in the plasma dissipate into heat \citep{rudakovMHDTurbulenceRadiating1997a,krouppTurbulentStagnationPinch2018-fixed}.
Therefore, to bridge the gap between observed, intermittently driven turbulent systems and theories of stationary MHD turbulence, we can study the behavior of decaying magnetized turbulence in an idealized environment.

In decaying turbulence, the turbulent flow arises purely from the initial conditions in the absence of a continuous driving force that injects energy.
Essentially, the driving force is a delta function forcing at the initialization of the flow.
The absence of external forces can avoid some of the shortfalls of driven turbulence simulations.
As an example of these shortfalls, previous studies have shown that seemingly unimportant driving parameters such as the autocorrelation time and normalization of the driving field can bias plasma properties in turbulence simulations, in some cases affecting the scaling of the energy spectra \citep{greteMatterForceSystematic2018}.
In addition, the driving forces contaminate the driven scales, making studies of turbulent plasma properties on those scale difficult to interpret.
Simulations of decaying turbulence with fixed initial conditions avoid these issues since there are no driving forces.

The Taylor-Green (TG) vortex provides a useful set of smooth initial conditions that devolve into a turbulent flow.
It was first proposed by \citet{taylorMechanismProductionSmall1937} as an early mathematical exploration of the development of the turbulent cascade in a three dimensional hydrodynamic fluid.
In the modern era, it is a canonical transition-to-turbulence problem also used for validation and verification of numerical schemes \citep{wang2013high}.
From a physics point of view, the TG vortex has been explored from numerous angles, including numerical simulations of inviscid and viscous incompressible hydrodynamics with an emphasis on the development of small scale structures through vortex stretching \citep{brachetSmallscaleStructureTaylor1983}.
Multiple configurations for TG vortices with magnetic fields were proposed in \citet{leeParadigmaticFlowSmallscale2008} in order to study decaying turbulence in incompressible MHD.
The new magnetic field configurations maintain all of the symmetries of the original hydrodynamic flow \citep{leeParadigmaticFlowSmallscale2008}, and later works \citep{leeLackUniversalityDecaying2010,pouquetDynamicsUnforcedTurbulence2010,brachetIdealEvolutionMagnetohydrodynamic2013} used these symmetries to save computational resources and allow more highly resolved simulations of the vortex. These simulations produced differing $k^{-2}$, $k^{-5/3}$, and $k^{-3/2}$ spectra  depending on the initial magnetic field, where the $k^{-2}$ spectra was speculated to be due to weak turbulence.
Later work by \citet{dallasOriginsEnsuremath2Spectrum2013-fixed,dallasStructuresDynamicsSmall2013} investigated the mechanism behind the different spectra. They concluded that the $k^{-2}$ spectra produced by one configuration of the magnetic field was due to magnetic discontinuities in the plasma and not weak turbulence as previously thought. In \citet{dallasSymmetryBreakingDecaying2013}, perturbations added to the initial conditions lead the symmetries of the TG vortex to break and the $k^{-2}$ spectra to dissipate to shallower $k^{-5/3}$ spectra. 
A similar problem using the hydrodynamic initial configuration of the TG vortex but with an Orszag-Tang magnetic field was studied in imcompressible resistive MHD by \citet{vahalaMHDTurbulenceStudies2008-fixed}, where a $k^{-5/3}$ energy spectra was found in their simulations.

All of these studies are concerned with incompressible turbulence, whereas many astrophysical systems (such as the interstellar, circumgalactic, intracluster, and intergalactic media) are comprised of compressible magnetized plasmas.
To our knowledge, the formulation of the TG vortex from \citet{leeParadigmaticFlowSmallscale2008} remains unexplored in the compressible MHD regime.
Moreover, there have been recent advances in analytical tools to study the transfer of energy between reservoirs in compressible MHD \citep{Yang2016, grete_energy_2017}.
Energy transfer analysis enables measurement of the flux of energies between length scales within and between the kinetic, magnetic, and thermal energies of the plasma.
In a compressible ideal MHD plasma, energy can be redistributed within the kinetic and within the magnetic energy budget via advection and compression.
Moreover, magnetic tension can facilitate energy transfer between kinetic and magnetic energies as vortical motion in the turbulent plasma contributes to magnetic fields and magnetic fields constrain the motion of the plasma.  
In turbulent flow, intra-budget energy transfers via advection and compression typically manifest from a larger scale to a  smaller but similar scale (i.e., ``down scale-local"),
defining the turbulent cascade.
Inter-budget energy transfer via, e.g., magnetic tension, complicates the picture of a turbulent cascade as it moves energy between reservoirs and potentially allows for nonlocal transfer of energy from large scales directly to much smaller scales.
Given the transient nature of the TG vortex, we expect the energy transfers to change over time as, e.g., the ratio of kinetic to magnetic energy evolves over time or due to the development of increasingly small-scale structure.
This is in contrast to stationary turbulence where the dynamics remain constant over time in a statistical sense.

For these reasons, we focus on a detailed study of the dynamics in the magnetized, weakly compressible Taylor-Green vortex.
Moreover, to explore magnetized decaying turbulence in different regimes we present nine simulations of the TG vortex probing all combinations of three different initial ratios of kinetic to magnetic energy ($1$, $10$, and $100$, corresponding to initial Alfv\'enic Mach numbers of $\mathcal{M}_A = \{1, 3.2, 10\}$) and three different initial fluid velocities (initial root mean squared, or RMS, sonic Mach numbers of $\mathcal{M}_{s,0}=\{0.1,0.2,0.4\}$). 
Thus, we explore strongly and weakly magnetized, subsonic plasmas in which density perturbations are present but limited. 

To summarize our results, we find that magnetic fields significantly influence the decaying turbulence in the plasma regardless of the initial field strength. 
In all cases, we find that at late times the magnetic dynamics dominate kinetic dynamics even if the initial magnetic energy is 100 times smaller than the kinetic energy.
Moreover, the spectral indices of the kinetic and magnetic energies are not fixed in time but evolve from steep $\simeq k^{-2}$ spectra at earlier times to shallower $\simeq k^{-4/3}$ spectra at later times.
Using the energy transfer analysis, we see that most energy transfer is dominated by magnetic field dynamics.
This includes both energy flux from kinetic to magnetic energy via magnetic tension and the flux of energy within the magnetic energy budget via compression and advection.
Overall, the kinetic energy cascade is effectively absent and the initial sonic Mach number ($\mathcal{M}_{s,0}$) only weakly affects the observed dynamics.
We also see several transient phenomena during the transition to turbulence, including temporary inverse turbulent cascades in both the magnetic and kinetic energies and large nonlocal energy transfers between scales separated by up to two orders of magnitude from the kinetic to the magnetic energy.

We organize the paper as follows.
In Section~\ref{sec:method}, we describe the simulation and analysis setup including numerical methods, detailed Taylor-Green vortex initial conditions, and the energy transfer analysis.
In Section~\ref{sec:results}, we present results of the simulations (focusing on $\mathcal{M}_{s,0} = 0.2$) such as the bulk properties of the plasma, the evolution of the energy spectra, and the transient behaviors seen through the energy transfer analysis:
In Section~\ref{sec:discussion}, we discuss our findings in the broader context of magnetized turbulence and astrophysical plasmas and conclude in
Section~\ref{sec:conclusions} with a summary of our key findings.
The online supplementary materials for this paper contain detailed plots of the results of all initial $\mathcal{M}_{s,0}$.

\section{Method}
\label{sec:method}

\subsection{MHD Equations and Numerical Method}
\label{sec:mhd_equations}
The equations for compressible ideal MHD plasma can be written as a
hyperbolic system of conservation laws. In differential form the ideal MHD equations are
\begin{eqnarray*}
    \partial_t \rho + \nabla \cdot \left ( \rho \mathbf{u} \right )  = 0\\
    \partial_t \rho \mathbf{u} + \nabla \cdot \left ( \rho \mathbf{u} \otimes \mathbf{u} - \mathbf{B} \otimes \mathbf{B} \right ) +
    \nabla \left (p + \mathbf{B}^2/2 \right ) = 0 \\
    \partial_t \mathbf{B} - \nabla \times \left ( \mathbf{u} \times \mathbf{B} \right ) = 0 \\
    \partial_t E + \nabla \cdot \left [ 
    \left ( E + p + \mathbf{B}^2/2 \right ) \mathbf{u} - 
    \left( \mathbf{B} \cdot \mathbf{v} \right ) \mathbf{B} \right ] = 0
\end{eqnarray*}
where $\rho$ is the density, $\mathbf{u}$ is the flow velocity, $\mathbf{B}$ is the magnetic field (that includes a factor of $1/\sqrt{4 \pi}$), $p$ is the thermal pressure, and $E$ is the total energy density.
We close the system of equations with the equation of state for an adiabatic ideal gas with
\begin{eqnarray*}
p=\rho \left (\gamma -1 \right) e
\end{eqnarray*}
where $\gamma$ is the ratio of specific heats and $e$ is the internal energy 
found from
\begin{eqnarray*}
E=\rho \left ( \frac{1}{2}\mathbf{u}\cdot\mathbf{u} + \frac{1}{2}\mathbf{B}\cdot\mathbf{B} + e \right ).
\end{eqnarray*}

We use the open source \kathena \cite{greteKAthenaPerformancePortable2021} astrophysical MHD code, which is a performance portable version of \Athenapp \cite{stoneAthenaAdaptiveMesh2020} using the \kokkos performance portability library \cite{carteredwardsKokkosEnablingManycore2014}.
\kathena uses an unsplit finite volume Godunov scheme to evolve the ideal MHD equations originally presented and implemented in \Athena \cite{Stone2009}.
The method consists of a second-order Van Leer predictor-corrector integrator with piecewise linear reconstruction (PLM) and HLLD Riemann solver, and constrained transport to preserve a divergence-free magnetic field.

\subsection{Magnetized TG Vortex}
\label{sec:magnetized_taylor_green_vortex}

The TG vortex was first proposed by \citet{taylorMechanismProductionSmall1937} as a mathematical exploration of the development of hydrodynamic turbulence in 3D.
The initial flow was made to be periodic and symmetrical in order to accommodate simple approximations to a solution.
There exist a number of different formulations. We follow the setup described in \citet{wang2013high} for the hydro variables and \citet{leeParadigmaticFlowSmallscale2008} for the initial magnetic field configuration.

The simplest hydrodynamic setup of a TG vortex begins with a periodic field of fluid velocity in the xy-plane and periodic pressure and density field with constant sound speed throughout the domain.
Using a cubic periodic domain with side length $2 \pi L$, the initial fluid velocity is set to
\begin{eqnarray*}
u_x  &=&  u_0 \sin { \frac{x}{L} }\cos { \frac{y}{L} }\cos { \frac{z}{L} }  \\
u_y  &=& -u_0 \cos {  \frac{x}{L} } \sin { \frac{y}{L} }\cos { \frac{z}{L} } \\
u_z  &=& 0 
\end{eqnarray*}
where $u_0$ is the maximum initial velocity.
Note that in this formulation the initial flow velocity is confined to the xy-plane.
The initial pressure and density are set to 

\begin{eqnarray*}
P    &=& P_0 + \frac{\rho_0 u_0^2}{16} \left ( \cos  \frac{ 2 x }{L}  + \cos \frac{ 2 y}{ L} \right )
                                \left ( \cos \frac{2 z}{L} + 2 \right )\\
\rho &=& P \rho_0/P_0
\end{eqnarray*}

so that $P$ and $\rho$ are proportional to each other.
This means that the sound speed
\begin{eqnarray*}
c_s = \sqrt{\gamma P/\rho} = \sqrt{\gamma P_0/\rho_0} 
\end{eqnarray*}
is initially constant throughout the domain.

The root mean square (RMS) of the initial Mach number is related to $u_0$ by
\begin{eqnarray*}
\mathcal{M}_{s,0} = \frac{u_0}{2 c_s}.
\end{eqnarray*}

For simplicity, we set $P_0=1$ and $\rho_0=1$.  We assume the fluid is a monatomic ideal gas with an adiabatic index $\gamma = 5/3$. 
The resulting total initial kinetic energy is 

\begin{equation}
E_{U,0} =  \rho_0 u_0^2\left ( \pi L \right)^3 \;.
\end{equation}

Magnetic fields were first added to the TG vortex in \cite{leeParadigmaticFlowSmallscale2008} with the express constraint of preserving the same symmetries of the hydrodynamic flow.
Here, we follow the proposed insulating configuration so that currents are confined to $\pi L$ boxes, e.g., the cube $[0,\pi L]^3$ forms an insulating box.
The corresponding initial magnetic fields are given by
\begin{eqnarray*}
B_x  &=& B_0 \cos {\frac{x}{L}} \sin {\frac{y}{L}} \sin{\frac{z}{L}} \\
B_y  &=& B_0 \sin {\frac{x}{L}} \cos {\frac{y}{L}} \sin{\frac{z}{L}} \\
B_z  &=& -2 B_0 \sin{\frac{x}{L}} \sin{\frac{y}{L}} \cos{\frac{z}{L}}
\end{eqnarray*}
where $B_0$ is the initial magnetic field strength.
In practice, we initialize the magnetic field from the magnetic vector potential $\mathbf{A}$
\begin{eqnarray*}
A_x &=& -B_0 \sin \left ( \frac{x}{L} \right ) \cos \left ( \frac{y}{L} \right ) \cos \left ( \frac{z}{L} \right ) \\
A_y &=&  B_0 \cos \left ( \frac{x}{L} \right ) \sin \left ( \frac{y}{L} \right ) \cos \left ( \frac{z}{L} \right ) \\
A_z &=& 0
\end{eqnarray*}
using $\mathbf{B}=\nabla \times \mathbf{A}$.
This guarantees $\nabla \cdot \mathbf{B} = 0$ to machine precision in the initial conditions, which is then
preserved by the constrained transport algorithm throughout the simulation.
The total initial magnetic energy is 
\begin{equation}
E_{B,0} = 3 B_0^2 \left ( \pi L \right )^3
\end{equation}
so that the initial ratio of kinetic to magnetic energy is 

\begin{equation}
\label{eq:energy_ratio}
\frac{E_{U,0}}{E_{B,0}} = \frac{ \rho_0 u_0^2 }{3 B_0^2}.
\end{equation}

Since the magnetic field is zero is some regions of the domain, the Alfv\'enic Mach number $\mathcal{M}_A=u \sqrt{\rho}/B$ is also undefined in some regions. 
For this reason, we use a proxy based on the mean energies for the Alfv\'enic Mach number
\begin{equation}
\mathcal{M}_A:=\sqrt{\langle E_U \rangle/\langle E_B \rangle }
\end{equation}
throughout the rest of the paper.
We also adopt a similar proxy for the plasma $\beta$ (ratio of thermal to magnetic pressure)
\begin{equation}
\beta:=\frac{2}{\gamma}\frac{ \mathcal{M}_A^2}{\mathcal{M}_s^2} \;
\end{equation}
where $\mathcal{M}_s$ is the RMS of the sonic Mach number.

The hydrodynamic and magnetic initial conditions exhibit a number of symmetries that are maintained throughout the simulation.
In each of the three dimensions there are two planes across which the fluid is antisymmetric.
For our setup, these are planes through $x=0$ and $x=\pi L$; planes through $y=0$ and $y=\pi L$; and planes through $z=0$ and $z=\pi L$.
Additionally, the flow is rotationally symmetric through a rotation of $\pi$ around the two axes $x=z=\pi L/2$ and $x=z=\pi L/2$ and rotationally symmetric through a rotation of $\pi/2$ around the axis $x=y=\pi L /2$.
These symmetries are more thoroughly explored in \cite{leeParadigmaticFlowSmallscale2008}.

We explore the transition to magnetized turbulence and the following decay in different regimes with our simulation suite of TG vortices and focus on 
two parameters: the initial RMS Mach number using $\mathcal{M}_{s,0} = \{0.1,0.2,0.4\}$ and the initial ratio of kinetic to magnetic energy using $E_{U,0}/E_{B,0}=\{1,10,100\}$, or alternatively, the initial RMS Alfv\'enic Mach number $\mathcal{M}_{A,0}=\{1,3.2,10\}$.
We simulate all nine combinations of the three values of these two parameters.
Throughout the rest of the text, we use \Ms{X} to refer to simulations with $\mathcal{M}_{s,0}=X$ and \Ma{Y} to refer to simulations with $\mathcal{M}_{A,0}=Y$.

The initial magnetic field amplitude $B_0$ is obtained from Equation \ref{eq:energy_ratio} using given a specific value of $\mathcal{M}_{s,0}$ and $\mathcal{M}_{A,0}$.
All simulations employ a cubic $[-0.5,0.5]^3$ domain with periodic boundaries, with $L=\frac{1}{2\pi}$ to be consistent with the definition of the initial condition that is presented above.  We use a uniform Cartesian grid with $1{,}024^3$ cells.
The characteristic length scale of the initial vortices is $\pi L$, so that we define
\begin{eqnarray*}
T=\frac{\pi L}{u_0}
\end{eqnarray*}
as the dynamical time
\footnote{Note that other works such as \citet{wang2013high,pouquetDynamicsUnforcedTurbulence2010}
use a nondimensional time, $t^* = L/u_0$, in contrast to the dynamical time used here.}
In order to evolve the simulations for sufficient time to allow a turbulent flow to form and evolve, we run each simulation for $\approx6$ dynamical times.

In our results, we present all measurements of time in terms of the dynamical time $T$ and all measurements of wavenumber in terms of $1/L$. Unless otherwise noted, all other results are in terms of simulation units.

\subsection{Energy Transfer Analysis}
\label{sec:energy_transfer_analysis}

In order to probe the movement of energy between different energy reservoirs, we use the shell-to-shell energy transfer analysis from \citet{grete_energy_2017}, which extends the framework presented in \citet{Alexakis2005} to the compressible regime.

The total transfer of energy from some shell $Q$ in energy reservoir $X$ to some shell $K$ in reservoir $Y$ is denoted by 
\begin{equation}
\mathcal{T}_{XY} (Q,K) \quad X,Y \in [U,B]
\end{equation}
where we use $U$ and $B$ to denote the kinetic and magnetic energy reservoirs, respectively.

In this work we focus on the energy transfer within the kinetic and magnetic energy reservoirs via advection and compression which are respectively
\begin{eqnarray*}
\mathcal{T}_{UU} (Q,K)= & - \int \mathbf{w}^K \cdot \left ( \mathbf{u} \cdot \nabla \right ) \mathbf{w}^Q d \mathbf{x} \\
                        & - \frac{1}{2} \int \mathbf{w}^K \cdot \mathbf{w}^Q \nabla \cdot \mathbf{u} d \mathbf{x}\\
\mathcal{T}_{BB} (Q,K)= & - \int \mathbf{B}^K \cdot \left ( \mathbf{u} \cdot \nabla \right ) \mathbf{B}^Q d \mathbf{x} \\
                        & - \frac{1}{2} \int \mathbf{B}^K \cdot \mathbf{B}^Q \nabla \cdot \mathbf{u} d \mathbf{x}                    
\end{eqnarray*}
and the energy transferred from kinetic energy to magnetic energy via magnetic tension (and vice versa) given by
\begin{eqnarray}
\mathcal{T}_{UBT} (Q,K)&=& \int \mathbf{B}^K \cdot \nabla \left ( \mathbf{v}_A \otimes \mathbf{w}^Q \right ) d \mathbf{x} \\
\mathcal{T}_{BUT} (Q,K)&=& \int \mathbf{w}^K \cdot \left ( \mathbf{v}_A \cdot \nabla \right ) \mathbf{B}^Q d \mathbf{x} \;.
\end{eqnarray}
Here we use the mass weighted velocity $\mathbf{w}=\sqrt{\rho} \mathbf{u}$ so that the spectral energy density is positive definite \cite{kidaEnergySpectralDynamics1990}, and $\mathbf{v}_A$ is the Alfv\'enic wave speed.

The velocity $\mathbf{w}^K$ and magnetic field $\mathbf{B}^K$ in a shell K (or Q) are obtained using 
a sharp spectral filter in Fourier space.
The shell bounds are logarithmically spaced and given by $1$ and $2^{n/4 + 2}$ for $n \in \{ -1,0,1,\ldots,32\}$.
Shells (uppercase, e.g., K) and wavenumbers (lowercase, e.g., $k$) obey a direct mapping, i.e.,
$K=24$ corresponds to the logarithmic shell that contains $k = 24$, i.e., $k \in (22.6,26.9]$. 

\section{Results}
\label{sec:results}

In this section we present results of the Taylor-Green vortices we
simulated, showing bulk properties of the fluid
(Section~\ref{sec:bulkprops}), including the evolution of the different energy
spectra.  These results demonstrate that the kinetic, magnetic, and thermal
energy reservoirs interact with each other in a manner that depends
significantly on the initial strength of the magnetic field.  The energy spectra evolves to a turbulent cascade over 1-2 dynamical times and then stays there for the remainder of the simulation.
In  Section~\ref{sec:energy_transfer}, we examine in
detail the transfer of energy between different energy
reservoirs, including the
transient behaviors we observed in the simulations.  We see robust
transfer of energy at all scales within the kinetic and magnetic
energy reservoirs when examined separately, as well as complex and
time-varying nonlocal transfer of energy between the kinetic and
magnetic energy reservoirs, including evidence for an intermittent inverse
turbulent cascade.
Since the initial Mach number had much less of an effect on the results compared to the initial ratio of kinetic to magnetic energy, we focus on results using only the three \Ms{0.2} simulations as reference.
We provide more complete plots of all nine simulations spanning all Mach numbers in the online supplements.

\begin{figure}[ht]
\includegraphics[width=\linewidth]{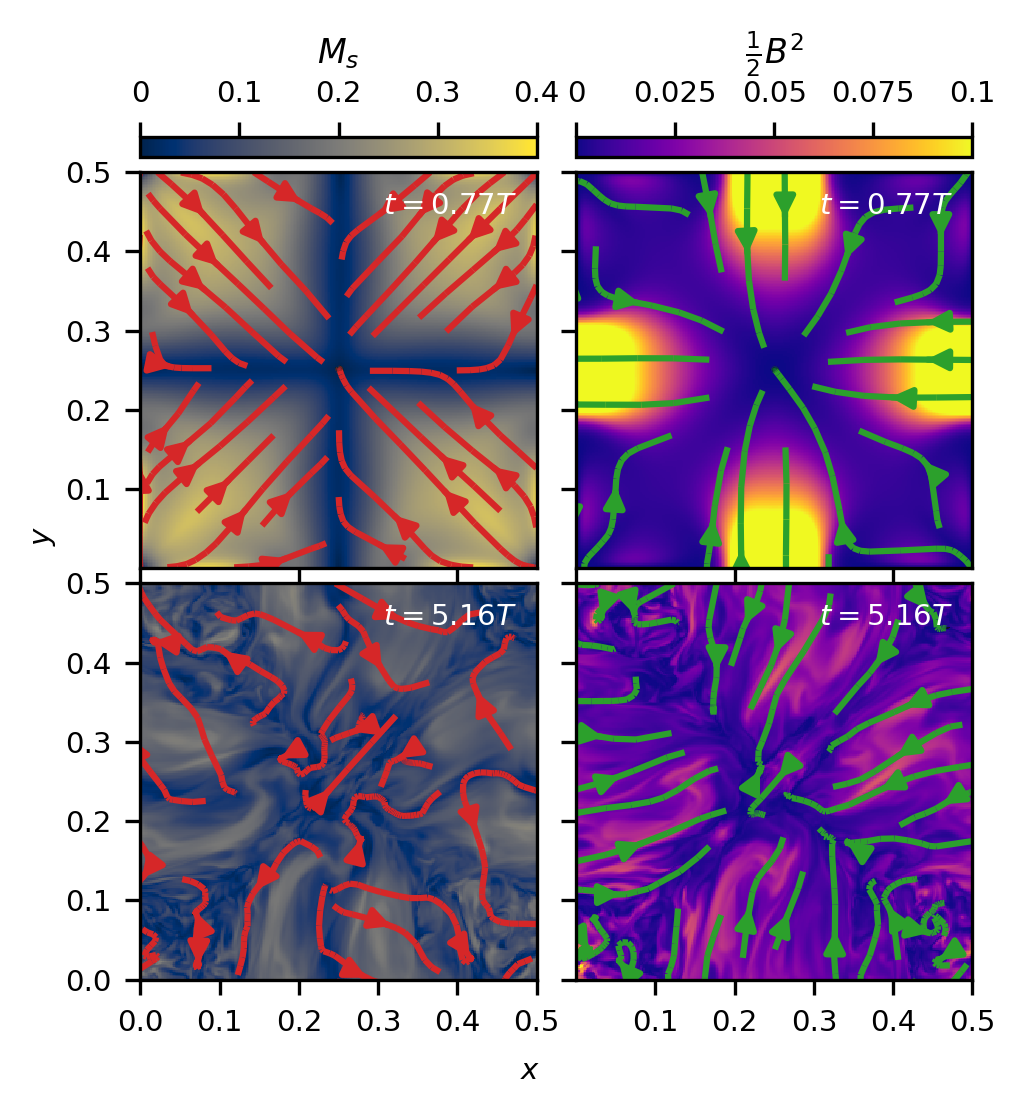}
\caption{\label{fig:slice_plot}
Slices of sonic Mach number (left) and magnetic pressure (right) at
$t=0.77T$ and $t=5.16T$ in the $xy-$plane through $z=\frac{\pi}{2}L$, with streamlines on the left showing the direction of flow and streamlines on the right showing the direction of the magnetic fields, plotting only the 1st quadrant from the \MsMa{0.2}{10} simulation, demonstrating the transition of the flow into turbulence.
}
\end{figure}
Starting with a visual demonstration of the TG vortex, Figure~\ref{fig:slice_plot} shows the sonic Mach number and magnetic pressure from the \MsMa{0.2}{10} simulation after $0.77$ dynamical times and after $5.16$ dynamical times in a slice in the $xy-$plane through the origin. 
Only one quadrant of the $xy$-place is shown, as it exhibits symmetry across 4 quadrants in the $xy$-plane. 
From the slice plot, we can see that the TG vortex begins as a smooth vortical flow and magnetic field. 
After several dynamical times, the smooth flow devolves into a chaotic magnetized turbulent flow.
Kinetic and magnetic structures at all scales persist throughout the simulation, as will be shown in energy spectra later in this work.

\subsection{Bulk Properties}
\label{sec:bulkprops}

\subsubsection{Evolution of energy reservoirs}
\label{sec:totalenergy}

\begin{figure*}[ht]
\includegraphics[width=\linewidth]{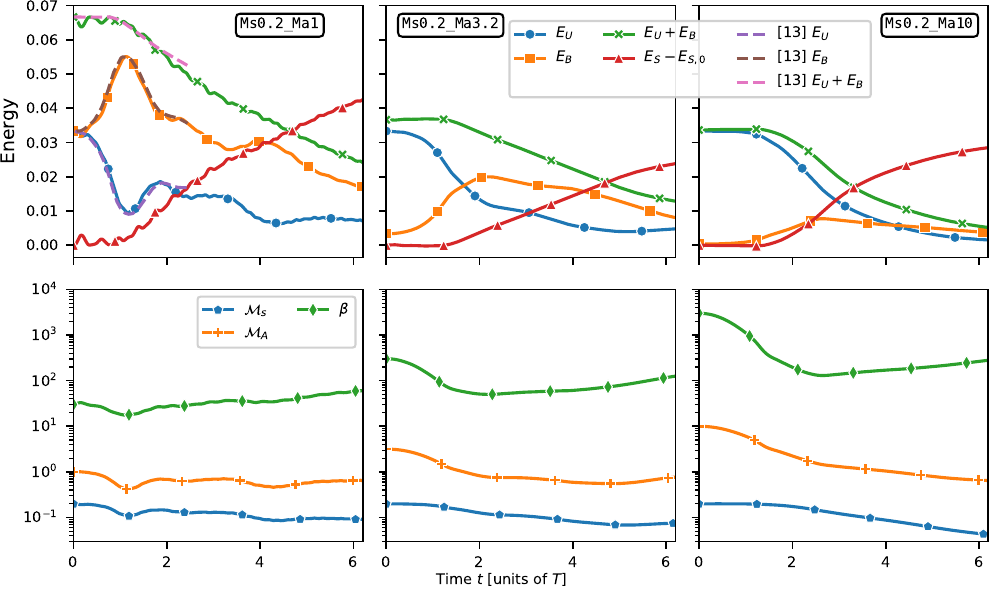}
\caption{\label{fig:energy_histories}
Mean energies over over time in the top row with kinetic energy (solid blue),
  magnetic energy (solid orange), the sum of kinetic and magnetic energies
  (solid green), and the change in thermal energy since the simulation start
  (solid red), and dimensionless numbers over time in the bottom row with RMS
  sonic Mach number $\mathcal{M}_s$ (blue), Alv\'enic Mach number
  $\mathcal{M}_A$ (orange), and plasma beta $\beta$ (green) for the \Ms{0.2}
  simulations. Energy over time from the simulation from Fig.~3a in
  \citet{pouquetDynamicsUnforcedTurbulence2010} (adjusted to the normalization
  used here), which matches the setup of the \MsMa{0.2}{1} simulation, is shown
  with dashed lines in the upper left panel for reference. 
  Energies and mach numbers for all nine simulations are shown in the online supplements.
  }
\end{figure*}

Figure \ref{fig:energy_histories} shows the total kinetic, magnetic, and
thermal energies and the dimensionless RMS sonic Mach number $\mathcal{M}_s$,
Alv\'enic Mach number $\mathcal{M}_A$, and plasma beta $\beta$  of the \Ms{0.2}
simulations as a function of time.  In this figure, we can see that in all
simulations kinetic and magnetic energy convert into thermal energy over time.
This decay into thermal energy is not immediate; rather, it requires at least
one dynamical time to begin (i.e., it is observed to occur at a minimum of
$t=1T$ in all simulations).  In the \Ma{1} simulations, due to the initial
conditions there is even a small transient transfer of thermal energy into
kinetic and magnetic energies.  After $t=2T$, all simulations dissipate kinetic
and magnetic energy into thermal energy.  The sonic Mach number generally
decreases by less than a factor of 4 over time from its initial $0.2$ value,
and $\beta$ remains high (from $\gtrsim 20$ for \MsMa{0.2}{1} to $\gtrsim100$
for \MsMa{0.2}{10}) throughout the simulations.

In all cases, the flow becomes dominated by magnetic energy (i.e., become sub-Alfv\'enic with $\mathcal{M}_A < 1$) at different dynamical times depending on the initial ratio of kinetic to magnetic energy and mostly independent of the initial Mach number.
In other words, even for the simulations with initially 100 times more kinetic than magnetic energy (\Ma{10}), in the final state the  magnetic energy dominates over the kinetic energy.  
This already highlights the importance of kinetic to magnetic energy transfer.
The initial growth of magnetic energy is characteristic of the insulating magnetic field configuration and is seen in other works on the TG vortex \cite{leeLackUniversalityDecaying2010}. 
This behavior of the magnetic field is likely due to the magnetic fields and vorticity beginning parallel to each other everywhere.
All simulations experience a peak in the magnetic energy evolution before $t=3T$
depending on the initial magnetic energy.  
At $t=6T$, all simulations are still losing total kinetic and magnetic energy to thermal energy, although the rate of energy dissipation is slowing by the simulation end. 
The magnetic and kinetic energies also become similar in magnitude, cf., $\mathcal{M}_A \simeq 1$.

The \MsMa{0.2}{1} simulation displays notably different
behavior than those where the kinetic energy initially dominates.
 In particular, we observe periodic exchanges of energy between these two
reservoirs before the bulk of the energy is converted into heat,
rather than a smooth transfer of energy from the kinetic to magnetic
reservoir, followed by a decline of both as the flow
thermalizes.
At approximately $t=1T$, more than five times as much energy is stored
in the magnetic reservoir as compared to the kinetic reservoir, which
is in stark contrast with other calculations.
These results suggest that the large initial magnetic field
facilitates a more rapid transfer of kinetic energy, which will be
examined in more detail later in this paper.
For reference, we also plot the temporal evolution of the energies in the 
incompressible, magnetized Taylor-Green vortex with $\Ma = 1$ presented in \citet{pouquetDynamicsUnforcedTurbulence2010}
in the top left panel of Fig.~\ref{fig:energy_histories} next to our \MsMa{0.2}{1} results.
The evolution in \citep{pouquetDynamicsUnforcedTurbulence2010} covers the first oscillation and is in 
good agreement with our simulation.
Finally, the oscillations observed in the energy reservoirs for the \Ma{1} simulations in general
have a period that depends on the initial Mach number, which can be seen in the figures that we leave for the online supplements.

\subsubsection{Energy Spectra}
\label{sec:energy_spectra}

\begin{figure*}[] 
\includegraphics[width=\linewidth]{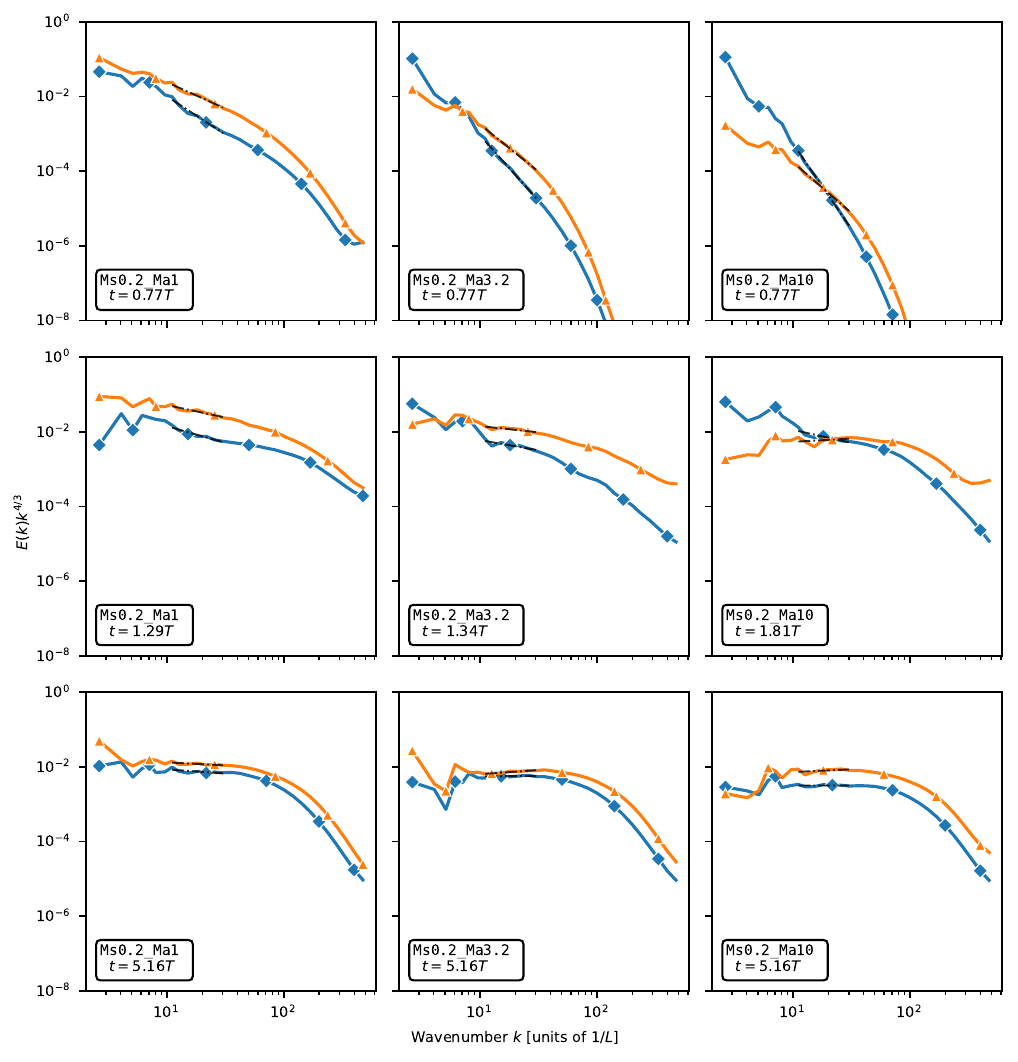}
\caption{\label{fig:spectra_over_time}
Kinetic energy spectra (in solid blue) and magnetic energy spectra  (in solid orange)
compensated by $k^{4/3}$, with black dashed lines showing the power law fit to the spectral to obtain a spectral index. 
In the left column we show the \MsMa{0.2}{1} simulation, in the middle column we show the \MsMa{0.2}{3.2} simulation, and in the right column we show the \MsMa{0.2}{10} simulation. 
In the top row we show all simulations at $t=0.77T$, in the middle row we show the three simulations at different times ($t=1.29$, $t=1.81T$, $t=1.81T$) when the simulations are displaying interesting behavior discussed in sections \ref{sec:inverted_turbulent_cascades} and \ref{sec:non_local_energy_transfer}, and in the bottom row we show all simulations at $t=5.16T$ when the initial flow has completely decayed into turbulence and both energy spectra fluctuate around a $k^{-4/3}$ spectrum.
}
\end{figure*}

Figure \ref{fig:spectra_over_time} shows the temporal evolution of the
kinetic and 
magnetic energy spectra of the three \Ms{0.2} simulations, compensated by $k^{4/3}$,
which demonstrates how both the kinetic and magnetic energy spectra change from the smooth 
initial large scale flow to fully developed turbulence.
The top row shows the three simulations earlier in the evolution ($t=0.77T$), when the spectra are still steep with large scale structure from the initial conditions.
In the case of the strongest initial magnetization (\Ma{1}), the magnetic energy is larger 
than the kinetic energy on all scales and their spectral scaling is comparable.
For \Ma{3.2} and \Ma{10} the kinetic energy spectrum is steeper than the magnetic one.
The spectra cross at $k\simeq7$ and $k\simeq20$, respectively, so that the kinetic energy
is still dominant on large scales.
The middle row in Figure~\ref{fig:spectra_over_time} shows intermediate times with \MsMa{0.2}{1} at $t=1.29T$, which is the time that is discussed in Section~\ref{sec:inverted_turbulent_cascades} 
and \MsMa{0.2}{3.2} and \MsMa{0.2}{10} simulations at $t=1.81T$, which is the time is discussed in Section \ref{sec:non_local_energy_transfer}.
Note that the spectra are still evolving at this intermediate stage.
In the \MsMa{0.2}{10} simulation at $t=1.81T$, the magnetic spectra has reached a $k^{-4/3}$ spectrum while the kinetic spectra shows a broken power law with excess energy at larger length scales.
In both \Ma{1} and \Ma{3.2} the magnetic energy is now dominant on effectively all scales
(with the exception of the noisy part of the spectrum at the largest scales, $k\lesssim 4$).
The bottom row shows all three \Ms{0.2} simulations at $t=5.16T$.
Here, the magnetic energy is effectively dominant on all scales in all simulations and the kinetic and magnetic spectra exhibit a scaling close to $k^{-4/3}$.
The spectral indices still fluctuate, which we explore in Section~\ref{sec:spectral_index}.

\begin{figure*}[] 
\includegraphics[width=\linewidth]{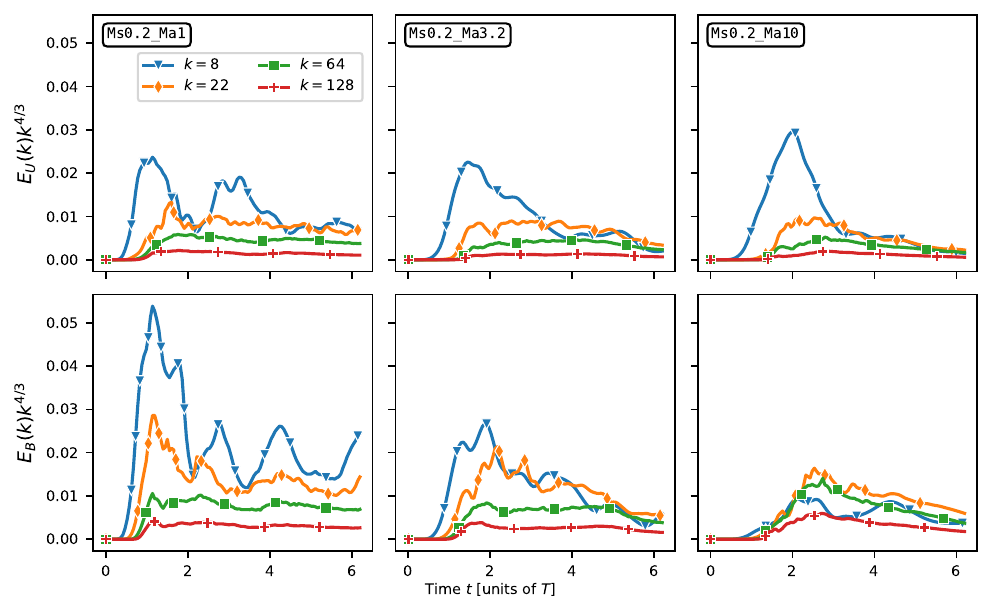}
\caption{\label{fig:energy_by_k}
The kinetic energy (top) and magnetic energy (bottom) at wavenumbers $k=8,22,64,128$ plotted separately in different colors versus time, where the energy at each wavenumber has been compensated by $k^{4/3}$ to make them comparable. 
In the left column we show the \MsMa{0.2}{1} simulation, in the middle column we show the \MsMa{0.2}{3.2} simulation, and in the right column we show the \MsMa{0.2}{10} simulation. Energy at the smallest length scales in both reservoirs saturates at $t\simeq1T$, $t\simeq1.5 T$, and $t\simeq2.5$ in the \MsMa{0.2}{1}, \MsMa{0.2}{3.2}, and \MsMa{0.2}{10} simulations respectively, showing approximately when the turbulence has developed at all scales.
}
\end{figure*}

In Figure \ref{fig:energy_by_k} we show the kinetic and magnetic energy at specific wavenumbers and compensated by $k^{4/3}$ plotted over time.
At early times (before $t= 2T$) the large scale ($k=8$) kinetic energy shows the fastest growth rate compared to smaller scales as expected from an initial entirely large scale configuration.
The kinetic energy at $k=8$ peaks between $t=1T$ and $t=2T$ with larger initial magnetic field leading to an earlier peak.
The magnetic energy at $k=8$ in the \MsMa{0.2}{1} simulation oscillates throughout the duration of the simulation, with the kinetic energy oscillating once.
No oscillatory behavior is observed in \MsMa{0.2}{3.2} and \MsMa{0.2}{10} for these quantities.
From this plot we can also see that the small scale ($k=128$) energies saturate at $t\simeq1T$, $t\simeq1.5 T$, and $t\simeq2.5 T$, respectively.

\subsubsection{Spectral Index}
\label{sec:spectral_index}

\begin{figure*}[ht]
\includegraphics[width=\linewidth]{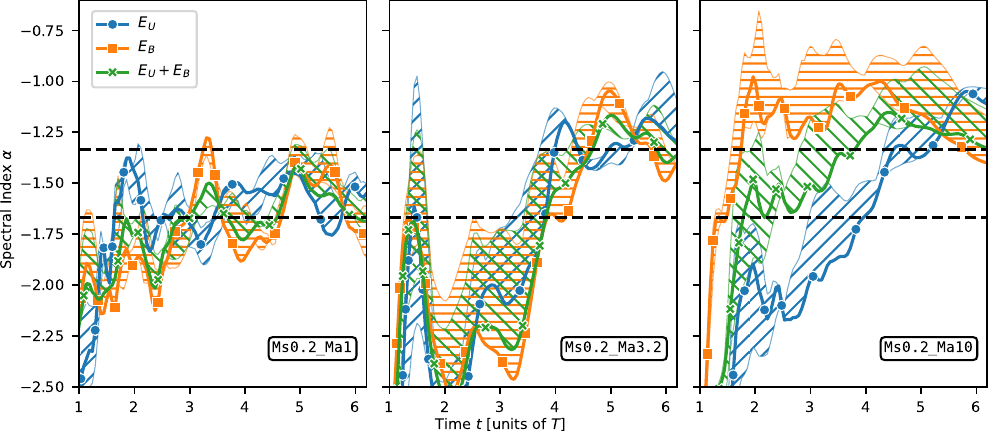}%
\caption{\label{fig:slopes_over_time}
Evolution of the spectral indices of the kinetic (blue), magnetic (orange), and sum of kinetic and magnetic energy (green) spectra over time for the \Ms{0.2} simulations.
The slope is computed from a least squares fit of the energy spectra limited to wavenumbers $k \in [10,32]$ which is approximately the inertial range. 
Shaded bands show how the fitted slope differs if a range $k \in [8,34]$, $k \in [10,32]$, or  $k \in [12,30]$ is used.
Note that the spectral index using the range $k \in [10,32]$ is not guaranteed to be bounded by the spectral indices obtained using $k \in [8,34]$, $k \in [10,32]$ and  $k \in [12,30]$, which is especially evident in the \MsMa{0.2}{3.2} and \MsMa{0.2}{10} simulations from $t\simeq 2T$ to $t\simeq 4T$. 
Horizontal dashed lines show $-4/3$ and $-5/3$ spectral indices.
The slope is only shown after $t=1T$ as the initial flow conditions dominate the spectra at early times, leading to steep spectra.
We include the spectral indices versus time for all nine simulations in the online supplements.}
\end{figure*}

We measured the spectral indices of the kinetic and magnetic energy spectra
$\alpha$ by fitting a power-law $E \propto k^{\alpha}$ to the energy spectra of
each reservoir at each time step.  For the inertial range of wavenumbers across
which we fit the power-law to the spectra, we used wavenumbers $k=10$ to
$k=32$.  We chose this inertial range because very little large scale structure
persists below $k=10$ and wavenumbers above $k=32$ are not entirely free of
numerical dissipation any more.  The  kinetic and magnetic spectral indices
measured across the inertial range are not fixed in time across the different
simulations, with the most variation being due to initial magnetic energy.
Figure \ref{fig:slopes_over_time} shows the spectral indices of the kinetic,
magnetic, and sum of kinetic and magnetic energy spectra over time for the
\Ms{0.2} simulations.  In all
simulations, the spectral index evolves over time, decaying from the initial
steep spectral index ($\alpha \lesssim -2$) as energy is transferred to small
scales.  The kinetic and magnetic spectral indices evolves separately in the
calculations until the magnetic energy exceeds the kinetic energy, after which
the spectral indices of the separate and combined reservoirs fluctuate within
$\Delta \alpha \simeq 0.2$.  The crossover of kinetic and magnetic energies
happens immediately  in the \MsMa{0.2}{1} simulation, early in the
\MsMa{0.2}{3.2} simulation before $t=2T$, and later in the \MsMa{0.2}{10}
simulation at $t\simeq4T$.  After the kinetic and magnetic spectral indices
reach rough parity and the magnetic field becomes dominant, both spectral
indices reach comparable values and reach a rough constant $1-2$ dynamical
times later, although they continue to vary over time.  Since the magnetic
fields in the \Ma{1} simulations immediately become dominant, the spectral
indices reach a rough constant at $t\simeq 2T$, while in the \Ma{3.2}
simulations they  reach a rough constant at $t\simeq 4T$ and in the \Ma{10}
simulations this happens at $t\simeq 5T$.  The \MsMa{0.2}{3.2} simulation
experiences a brief peak in the spectral index around $t\simeq 1.5 T$ while the
flow is still in transition. This is also reflected in the large uncertainty of
the spectral index during that time, e.g., the index of the kinetic energy
spectrum varies between $-1$ and $-2.25$ by choosing slightly different fitting
ranges (as indicated by the shaded blue bands in
Fig.~\ref{fig:slopes_over_time}).  Note that in the \Ma{10} case, the magnetic
spectrum flattens and the spectral index reaches a roughly constant value much
sooner than in the other two cases, at $t\simeq 2T$ when the kinetic energy
still dominates.  Later on in the \Ma{10} simulations, the kinetic spectral
index becomes comparable to  the magnetic spectral index.  For the high initial
magnetic field simulations, the spectral index levels out at about $\alpha
\simeq -5/3$ while the initially kinetically dominated simulations level out at
$\alpha \simeq -4/3$.

The final spectral indices depend on the initial ratio of kinetic to magnetic energy, with more magnetic energy leading to shallower magnetic spectra. 
The \Ma{1} simulations end with $\alpha \simeq -1.7$ (close to $-5/3$), \Ma{3.2} ends with $\alpha \simeq -1.3$ (close to $-4/3$), and \Ma{10} ends with slightly lower values of $\alpha \simeq -1.2$.
In the presence of the stronger magnetic fields in the \Ma{1} simulations, the flattening of the spectra seems to be suppressed.
Before the kinetic and magnetic spectral indices become comparable in each simulation, there is also greater variance in the spectral slope when measured using different inertial ranges.
This indicates that a power-law might be a poor fit for the spectra at those early times, showing that the spectra is not fully developed until the magnetic energy is dominant.
For example, as seen in Figure \ref{fig:spectra_over_time}, the kinetic energy spectra appears as a broken power law at intermediate times, which is especially evident in the \MsMa{0.2}{10} simulation at $t=1.81T$ to a lesser extent the \MsMa{0.2}{1} simulation at $t=1.29T$ and the \MsMa{0.2}{3.2} simulation at $t=1.81T$.
Oscillations in the spectral index of the \Ma{1} simulations also appear, whose period seems to be linked to the initial Mach number, with larger Mach numbers leading to a smaller period of oscillation.

We note that between the three values of $\mathcal{M}_A$, the simulations shown here exhibit a wide variety of behaviors, highlighted by the spectral indices in Fig.~\ref{fig:slopes_over_time}. More simulations with intermediate values of $\mathcal{M}_A$ would be required to determine if the transition between these behaviors is smooth or abrupt.

\subsection{Energy Transfer}
\label{sec:energy_transfer}

While the total energy and spectra of the kinetic and magnetic
reservoirs can broadly describe the isolated behavior of the different
energy reservoirs, examining the energy transfer within and between
reservoirs using the analysis described in Section
\ref{sec:energy_transfer_analysis} can provide deeper insights into
the physical phenomena, including demonstrating the mechanisms that
are responsible for the transfer of energy.
The shell-to-shell energy transfer fluxes examined in this section
demonstrate the flux from wavenumber $Q$ to wavenumber $K$ within and between
energy reservoirs via different pathways.

\begin{figure*}[ht]
\includegraphics[width=\linewidth]{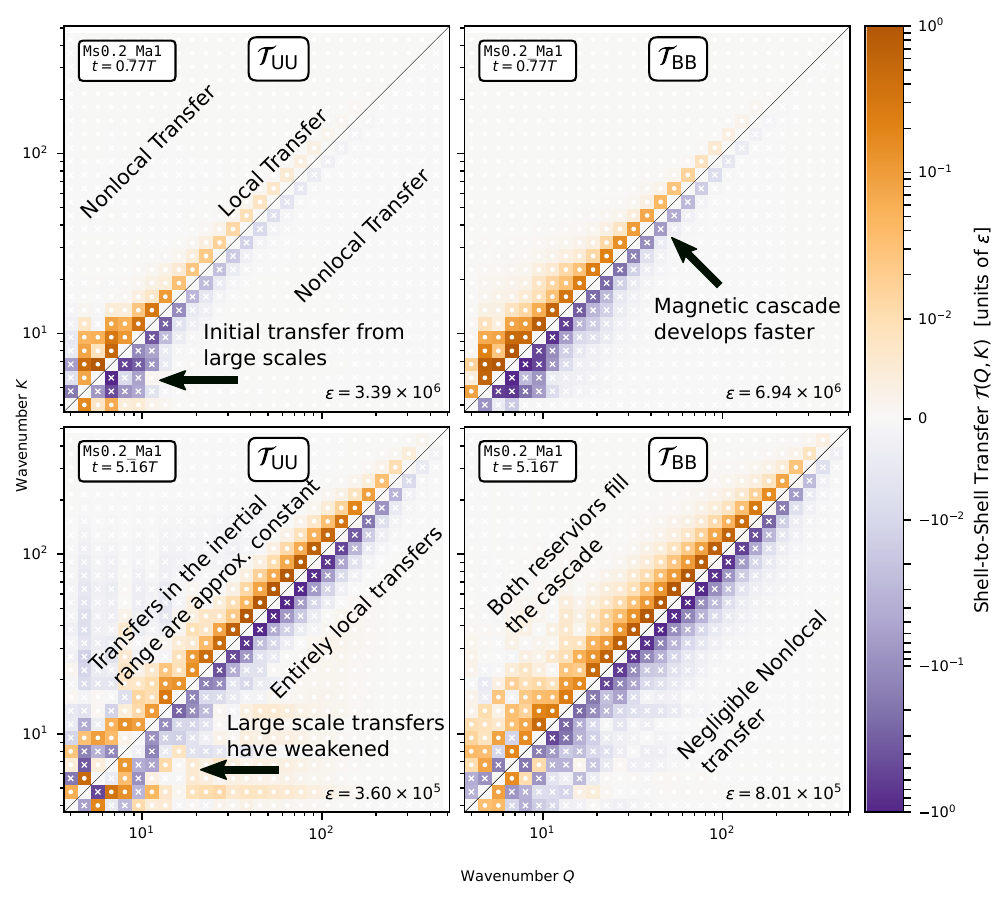}%
\caption{\label{fig:shell_to_shell_cascade}
Shell-to-shell energy transfer plots for the energy transfer within the kinetic (left) and magnetic (right) energy reservoirs  via advection and compression at $t=0.77 T$ (top) and $t=5.16 T$ (bottom) from the simulations with  \MsMa{0.2}{1}, showing the development of the kinetic and magnetic turbulent cascades. Annotations on the figure highlight key features of the energy transfer that are characteristic of a developing turbulence cascade.
Each bin shows the flux of energy from shell $Q$ to shell $K$, where orange with white circles showing a positive flux of energy, so that $K$ is gaining energy, and purple with white x's showing a negative flux, so that $K$ is losing energy. 
The energy flux in each bin is normalized by $\varepsilon = \max_{Q,K} |\mathcal{T}_{XY}(Q,K)|$ so that a higher $\varepsilon$ means a higher energy flux.
The solid black line shows equivalent scale transfers.
As the turbulent cascade develops in the magnetic and kinetic energy
reservoirs, more energy transfers along the diagonal fill out the
energy spectrum down to numerical dissipation scales.
}
\end{figure*}

Figure \ref{fig:shell_to_shell_cascade} shows the energy transfer
\textit{within} the kinetic (left) and magnetic (right) energy
reservoirs via advection
and compression in the \MsMa{0.2}{1} simulation at $t=0.77 T$ (top) and at $t=5.16 T$ (bottom).
This plot encapsulates the energy transfer of a turbulent cascade.
Near the beginning of the simulation in the top panels, most of the energy is in large scale modes, with energy from larger $Q$ wavenumbers moving to smaller $K$ wavenumbers.
Note that the energy transfer is constrained to the diagonal because the bulk of the energy transfer is local, occurring between comparable scales of $Q$ to $K$. 
White space fills the off-diagonals because very little nonlocal energy transfer occurs internally within reservoirs.
The energy transfer shown in this figure is solely within the kinetic and
magnetic reservoirs -- there is no energy transfer shown
between these reservoirs (although it is occurring, as will be
discussed in the next paragraph).
In the simulation shown here, the magnetic energy transfer is larger in magnitude than the kinetic energy transfer. In all simulations, the magnetic energy transfer extends to higher wavenumbers more rapidly than the kinetic energy. 
After the flow has decayed into turbulence (as shown in the bottom panels), energy transfer to smaller local scales happens across the resolved modes down to numerical dissipation scales. 
At large wavenumbers ($Q > 16$), the energy transfers are scale-local and of comparable magnitude.  This phenomenon continues to at least $Q \simeq 200$ in both the kinetic and magnetic energy transfer -- i.e., to much larger wavenumbers than an inertial range is observed (see, e.g., Figure~\ref{fig:spectra_over_time}).
Thus, the effective (numerical) viscosity and resistivity are not affecting the turbulent cascade encoded by these transfers to a significant degree.

\begin{figure}[ht]
\includegraphics[width=\linewidth]{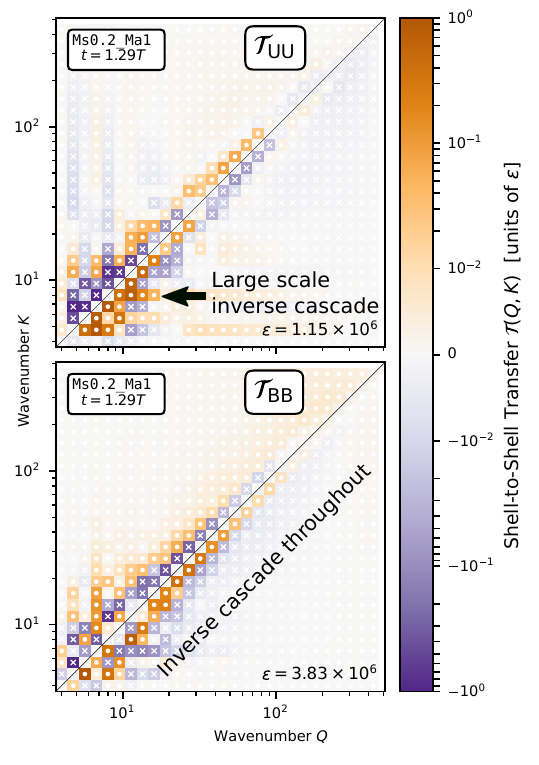}%
\caption{\label{fig:shell_to_shell_inverse_cascade}
Shell-to-shell energy transfer plots for the energy transfer within the kinetic (top) and magnetic (bottom) energy reservoirs  via advection and compression at $t=1.29 T$ from the \MsMa{0.2}{1} simulation, showing a transient inverse cascade within the magnetic energy reservoir (on all scales $K,Q \lesssim 100$) and kinetic energy reservoir (on large scales $K,Q \lesssim 16$). Annotations show where along the diagonal the inverse cascade is present.
}

\end{figure}
Figure \ref{fig:shell_to_shell_inverse_cascade} shows the energy transfer within the kinetic (top) and magnetic (bottom) energy reservoirs in the \MsMa{0.2}{1} simulation at $t=1.29 T$ (just before the magnetic energy peaks).
Energy transfer within the kinetic and magnetic reservoirs briefly reverses directions and moves energy from smaller local scales to larger local scales (note the purple color indicating energy loss above the diagonal and orange color below the diagonal, which is in contrast to Fig.~\ref{fig:shell_to_shell_cascade}).
This constitutes a transient inverse cascade.
Additionally, the inverse cascade is present throughout most scales of the magnetic energy ($K,Q \lesssim 100$) but only apparent at large scales in the kinetic energy ($K,Q \lesssim 16$).
As seen in Figure \ref{fig:energy_by_k}, at this early time the turbulent flow is just beginning to saturate the smallest scales while the large scale energy oscillates,
so the energy transfer inversion lasts less than a dynamical time (see Section \ref{sec:inverted_turbulent_cascades} for further exploration of the duration).

\begin{figure}[ht]
\includegraphics[width=\linewidth]{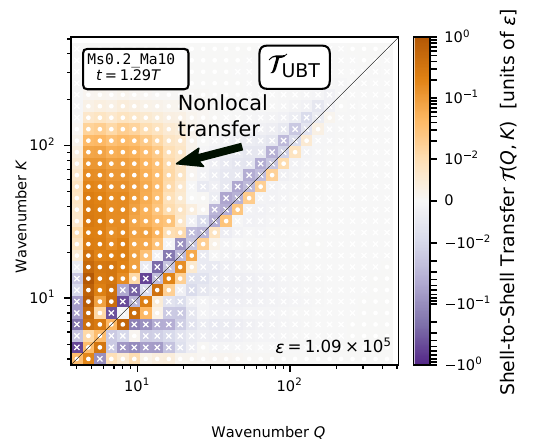}%
\caption{\label{fig:shell_to_shell_nonlocal}
Shell-to-shell energy transfer plots for the energy transfer from kinetic to magnetic energy via magnetic tension at $t=1.81 T$ from the \MsMa{0.2}{10} simulation, showing the nonlocal energy transfer from large kinetic scales to many smaller magnetic scales. Annotations show where the nonlocal transfer is present.
}
\end{figure}

Figure \ref{fig:shell_to_shell_nonlocal} shows the energy transfer
\textit{between} the kinetic to magnetic energy reservoirs due to
magnetic tension at $t=1.81T$  in the \MsMa{0.2}{10} simulation.
This Figure displays nonlocal transfer from kinetic to magnetic energy.
Unlike the advection- and compression-driven modes within the magnetic and kinetic energy reservoirs,
energy transfers from kinetic to magnetic reservoirs via tension can support nonlocal energy transfers.
The nonlocal transfer happens from large kinetic scales to much
smaller magnetic scales, spanning more than an order of magnitude downward in spatial
scale from the largest kinetic modes.
The nonlocal energy transfer between kinetic and magnetic energy was
significant in simulations with lower initial magnetic energy, and
especially in the \Ma{10} simulations where the magnetic field is
dynamically unimportant at early times.
Kinetic energy moves significant energy to all magnetic scales from early times at $t\simeq1.5T$ to
intermediate times at $t\simeq4T$ in these simulations, although some energy continues to flow via this mechanism at later times.
Additionally, since the transfer of energy via tension is between two
different reservoirs, the energy transfer can transfer at equivalent
scales from one reservoir to the other.  This is shown as non-zero
transfer along the diagonal of the plot. 

\subsubsection{Nonlocal Energy Transfer}
\label{sec:non_local_energy_transfer}

\begin{figure*}[ht]
\includegraphics[width=\linewidth]{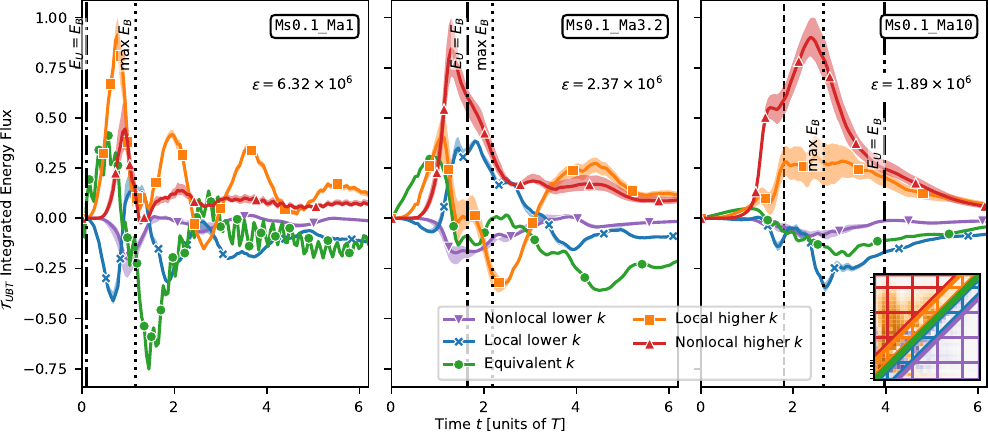}%
\caption{\label{fig:nonlocal_transfer}
Integrated energy flux over time from kinetic to magnetic energy via tension from larger wavenumbers to smaller nonlocal wavenumbers (purple),
from larger wavenumbers to smaller local wavenumbers (blue), between equivalent wavenumbers (green),
from smaller wavenumbers to larger local wavenumbers (orange), and from smaller wavenumbers to larger nonlocal wavenumbers (red) in the \Ms{0.2} simulations. 
We normalize the energy flux in each panel so that the absolute maximum of all of the flux bins is $1.0$, where $\varepsilon$ is the normalization factor use in each panel.
Comparisons of the relative strength of energy fluxes in different simulations must consider $\varepsilon$.
The inset plot in the lower right panel shows the color coded regions that are integrated to calculate each line at a single time for the same shell-to-shell transfer from Figure \ref{fig:shell_to_shell_nonlocal}.
Solid lines show the integrated flux if  ``local" wavenumbers as defined as 5 logarithmic bins away from the equivalent wavenumber. The shaded regions show the integrated flux if 4 or 6 bins are used, showing that the behavior is robust if the range ``local" wavenumbers is defined closer or further away from transfer between equivalent scales.
We include the integrated flux from kinetic to magnetic energy via tension for all nine simulations in the online supplements
}
\end{figure*}

Like in some driven turbulence simulations \cite{Alexakis2005,grete_energy_2017}, these decaying turbulence simulations also demonstrate significant nonlocal energy transfer between kinetic and magnetic energy reservoirs. 
Unlike in driven simulations, the energy transfers in this work are solely due to the fluid flow and not due to externally-applied driving forces.
Figure \ref{fig:nonlocal_transfer} shows the total local, nonlocal, and equivalent-scale energy transfers via magnetic tension in the \Ms{0.2} simulations over time.
We obtain these quantities by integrating the transfer functions over different sets of scales: 
\begin{eqnarray*}
&\text{Nonlocal lower}  & \sum_{Q}\sum_{K\in [1,2^{-\ell}Q )} \mathcal{T}_{XY}\left(Q,K\right)\\
&\text{Local-Lower}      & \sum_{Q}\sum_{K\in[2^{-\ell} Q,Q)} \mathcal{T}_{XY}\left(Q,K\right)\\
&\text{Equivalent}       &\sum_{Q}\sum_{K=Q} \mathcal{T}_{XY}\left(Q,K\right)\\
&\text{Local-Higher}     &\sum_{Q}\sum_{K\in(2^{\ell} Q,Q ]} \mathcal{T}_{XY}\left(Q,K\right) \\
&\text{Nonlocal Higher} &\sum_{Q}\sum_{K\in(2^{\ell} Q ,\infty]} \mathcal{T}_{XY}\left(Q,K\right)
\end{eqnarray*}
where $\ell$ is a parameter for differentiating local versus nonlocal separation of wavenumbers in log space. In Figure~\ref{fig:nonlocal_transfer}, we show the analysis using $\ell=5/4$ with a solid line, which corresponds to $5$ logarithmic bins above or below $Q$ (see \ref{sec:energy_transfer_analysis} for the description of the binning), and show the extent of the fluxes if $\ell=5/4\pm1/4$ is used in shaded regions.
As seen in this figure from the red line, the nonlocal energy transfer from large scale kinetic modes to small scale magnetic modes (``downscale'' transfer) is present in all simulations but is only dominant when the initial kinetic energy exceeds the initial magnetic energy -- this nonlocal energy transfer is more significant in the \Ma{3.2} and \Ma{10} simulations. 
Nonlocal energy transfer downscale (red line) peaks depending on the initial magnetic field and in all cases before the total magnetic energy peaks.
The nonlocal transfer helps fill out the magnetic energy spectrum faster than the kinetic energy spectrum, especially in the \Ma{10} simulations, which is consistent with the spectral index shown in Figure \ref{fig:spectra_over_time} and the turbulent cascades shown in the shell-to-shell energy transfer in Figure \ref{fig:shell_to_shell_cascade}.
By the time the magnetic energy has exceeded the kinetic energy in the \Ma{3.2} and \Ma{10} simulations, nonlocal energy transfer is largely diminished due to the lack of kinetic energy to feed the transfer.

Local energy transfer downscale (orange line) depends more strongly on the initial magnetic field, with local transfer to smaller scales reaching double the nonlocal transfer in the \Ma{1} simulation and being less than half in other cases.
Local energy transfer upscale (blue line) is positive for some early times in the \Ma{1} and \Ma{3.2} simulations.

The \Ma{1} simulations also display two different oscillatory 
behaviors, with a low frequency oscillation in the local energy transfer and a high frequency oscillation clearly visible in the equivalent energy transfer but also present in local and nonlocal down scale transfer.

\subsubsection{Inverted Turbulent Cascades}
\label{sec:inverted_turbulent_cascades}
\begin{figure}[ht]
\includegraphics[width=\linewidth]{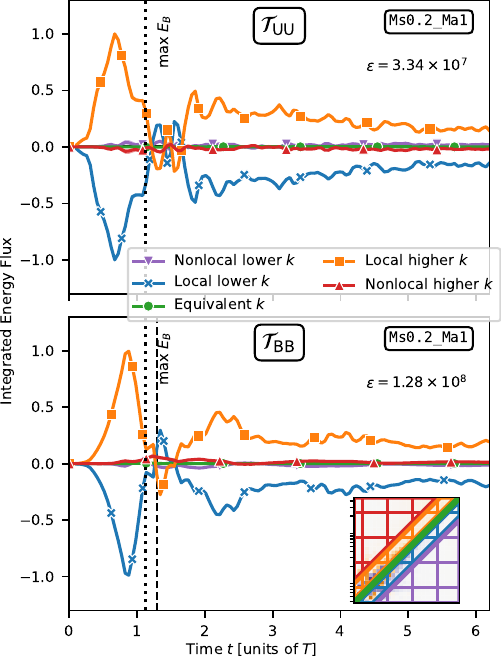}%
\caption{\label{fig:inverse_cascade}
Integrated energy flux over time within the kinetic energy (top) and within the magnetic energy (bottom) from larger wavenumbers to smaller nonlocal wavenumbers (purple),
from larger wavenumbers to smaller local wavenumbers (blue), between equivalent wavenumbers (green),
from smaller wavenumbers to larger local wavenumbers (orange), and from smaller wavenumbers to larger nonlocal wavenumbers (red) in the \MsMa{0.2}{1} simulation.
The inset plot in the lower middle panel demonstrates the color coded regions that are integrated to calculate each line at $t=1.29T$ from the shell-to-shell transfer from Figure~\ref{fig:shell_to_shell_inverse_cascade}.
Solid lines show the integrated flux if  "local" wavenumbers as defined as 5 logarithmic bins away from the equivalent wavenumber. The results change very little if 4 or 6 bins are used.
We include the integrated flux within the kinetic energy and magnetic energy for all nine simulations in the online supplements.
}
\end{figure}

At early times during the evolution of the \Ma{1} simulations, a temporary inverse cascade forms within the kinetic and magnetic energy reservoirs where small scale energy transfers to larger spatial scales.
Figure \ref{fig:inverse_cascade} shows the local and nonlocal energy transfers within the kinetic and magnetic energies to both smaller and larger length scales. 
In the \Ma{1} simulations, the local energy transfer from larger to smaller length scales temporarily reverses into an inverse cascade in both the kinetic and magnetic energy reservoirs shortly after peak magnetic energy is reached.  
The inversion appears with all three sonic Mach numbers simulated, with the longest inversion appearing in the \MsMa{0.1}{1} simulation for $\simeq 1T$ and shortest in the high \MsMa{0.4}{1} simulation for $\simeq 0.5T$.
For the \MsMa{0.1}{1} simulation, the  kinetic energy reservoir briefly reverses to the normal configuration, moving energy from large scales to scales while the magnetic energy is in an inverted cascade, before returning to the inverted cascade, lingering longer than the magnetic field in the inverted state and finally transitioning into a turbulent cascade for the rest of the simulation.
As seen in Figure \ref{fig:shell_to_shell_inverse_cascade}, the movement of energy to larger scales is not limited to any region of the spectra -- it is present at all length scales.
The \Ma{1} simulations, which are the only simulations to exhibit an inverse cascade, are also the only ones in which the total kinetic energy increases during any period.
After peak magnetic energy in the \Ma{1}, the magnetic energy increases while the kinetic energy increases for $\simeq1T$;
the inverse cascade appears during this same period.

\subsubsection{Cross-Scale Flux}
\label{sec:cross_scale}

\begin{figure*}[ht] 
\includegraphics[width=\linewidth]{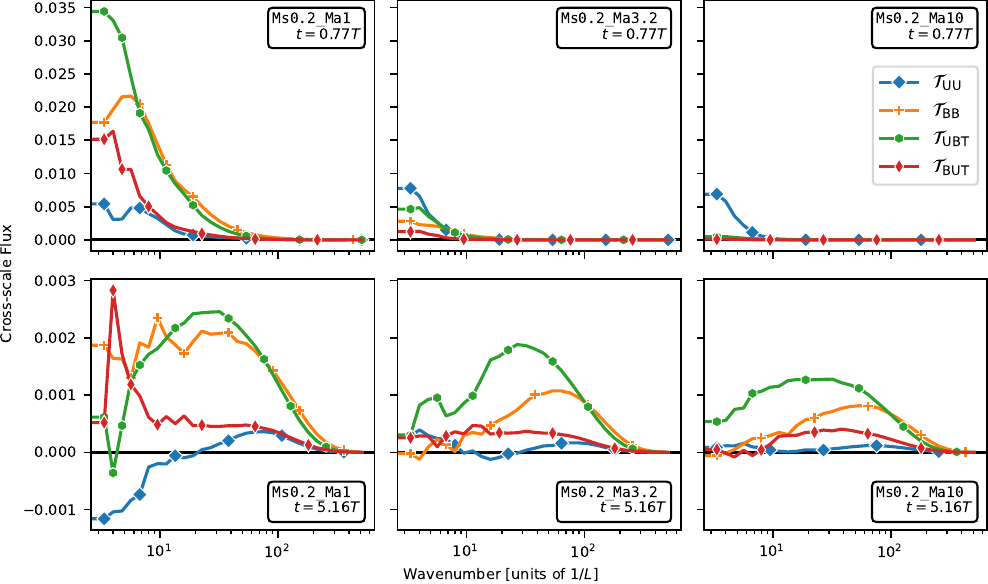}
\caption{\label{fig:cross_scale}
Cross-scale flux within the kinetic energy (blue line), within the magnetic energy (orange line), and from kinetic to magnetic energy via tension (green line) in the three \Ms{0.2} simulations across columns and at dynamical time $t=0.77 T$ (top) and later at dynamical time $t=5.16 T$. Note that the cross-scale fluxes at later times are an order of magnitude less than early cross-scale fluxes.  Positive values of this quantity denote energy transfer from larger to smaller scales.
}
\end{figure*}

With additional analysis of the shell-to-shell transfer, we can extract more insight into the movement of energy.
We can measure the cross-scale flux of energy from scales below a wavenumber $k$ to scales above a wave number $k$ by integrating the transfer function
\begin{equation}
    \Pi_{Y>}^{X<} (k) = \sum_{Q \leq k} \sum_{K\geq k} \mathcal{T}_{XY}\left (Q, K \right )
\end{equation}
Figure \ref{fig:cross_scale} shows the cross-scale fluxes via different transfer mechanisms for the simulations with \Ms{0.2}.
The top row shows cross-scale fluxes early in the simulation at $t=0.77 T$, when the large scale flow is still decaying into smaller scales.
The magnetic cross-scale flux at low wavenumbers predictably depends on the initial magnetic energy, while the kinetic energy cross-scale flux is largely the same between simulations at a given sonic Mach number.
For example, for \Ma{10} the cross-scale flux is strongly dominated by $\Pi_{U>}^{U<}$, whereas for \Ma{3.2} it is still the most significant contribution to the cross-scale flux, but substantial contributions are also seen from $\Pi_{B>}^{U<}$ ($\simeq 60\%$ of $\Pi_{U>}^{U<}(4)$), $\Pi_{B>}^{B<}$ ($\simeq 30\%$), and $\Pi_{U>}^{B<}$ ($\simeq 20\%$).
For the strongest initial magnetization ($\Ma{1}$) the early cross-scale flux is dominated by magnetic tension-mediated transfers from the kinetic-to-magnetic budget ($\Pi_{B>}^{U<}$) on all scales having a non-zero cross-scale flux ($k\lesssim 64$), with a similar contribution by the magnetic cascade on intermediate scales ($9 \lesssim k \lesssim 64 $).
The kinetic cascade is suppressed on all scales, generally contributing less than $10\%$ to the total cross-scale flux.

At later times ($t=5.16 T$, bottom row of Fig.~\ref{fig:cross_scale}), magnetic energy dominates both the energy budget and cross-scale energy flux.
Cross-scale energy flux via kinetic interactions is near zero across the inertial range of the spectrum, and thus does not significantly contribute to the total cross-scale energy flux. 
Only the magnetic fields facilitate down scale cross-scale flux at intermediate scales, both within the magnetic energy and from kinetic to magnetic energy.
Moreover, the relative contributions of the individual transfer $\Pi_{B>}^{U<}$, $\Pi_{B>}^{B<}$, $\Pi_{U>}^{B<}$, and $\Pi_{U>}^{U<}$ (in order of decreasing contribution) on intermediate scales ($16 \lesssim k \lesssim 64$) is the same independent of initial magnetization.
This continuous cross-scale flux is consistent with the evolving spectral index discussed in Section \ref{sec:spectral_index}.
Cross-scale flux through large physical scales is irregular, variable, and sometimes negative due to the lack of structure and driving forces at large scales. 

\section{Discussion}
\label{sec:discussion}

\subsection{Comparison to driven turbulence simulations}

The Taylor-Green vortex provides an interesting study of a freely evolving transition to decaying turbulence.
In other words, no external force is applied to the simulation as is the case in driven turbulence simulations.
This external force may introduce unintended dynamics to the flow \citep{greteMatterForceSystematic2018}.
For example, in a simulation that is mechanically driven at large scales, energy may still be injected on intermediate scales both in the incompressible regime \citep{Domaradzki2010} as
well as in the compressible regime due to density coupling \citep{grete_energy_2017}.
Moreover, mechanical driving generally results in an excess of energy on the excited, kinetic scales that
presents a barrier for magnetic field amplification on those scales in cases without a
dynamically relevant mean magnetic field.
This barrier is often expressed in the lack of a clear power law regime in the magnetic
spectrum and resembles an inverse parabolic shape.
At the same time, the magnetic energy spectrum drops below the kinetic one on the driving scales (see, e.g., Figure~1 in \citep{Grete2020slope} and references therein).
In the simulations presented here no such barrier is observed.
\label{sec:magnetic_spectral_index}
Both kinetic and magnetic energy spectra exhibit a (limited) regime where power law scaling is observed once a state of developed turbulence is reached.

Another important question raised from driven turbulence simulations pertains the locality of
energy transfers.
While there is agreement that $\mathcal{T}_{UU}$ and $\mathcal{T}_{BB}$ mediated transfers, i.e., within a budget, are highly local, the energy transfers between budgets (here, $\mathcal{T}_{UBT}$) have been observed to be weakly local and/or contain a nonlocal component from the driven scales \citep{Alexakis2005,Yang2016,grete_energy_2017}.
Here, we show that in the absence of the driving force the energy transfer mediated by magnetic tension contains both a local component as well as nonlocal component.
The latter directly transfers large-scale kinetic energy to large and intermediate scales in the magnetic energy budget.
Thus, the nonlocal component is not an artifact of an external driving force.

Finally, we recently showed that the kinetic energy spectra in driven turbulence simulations
follow a scaling close to $k^{-4/3}$, i.e., shallower than Kolmogorov scaling, and explained
this by the suppression of the kinetic energy cascade due to magnetic tension \citep{Grete2020slope}.
This is in agreement with our findings in the work presented here, where the same dynamics are observed at late times 
when turbulence is fully developed.

Naturally, this does not demonstrate that the same physical mechanisms are causing the similar slopes.
Nevertheless, the late time evolution of the simulations presented here is still comparable to a limited
degree to driven simulation of stationary turbulence.
For example, even at late times (see, e.g., $t=5.16T$ in Fig.~\ref{fig:shell_to_shell_cascade}), energy is still
cascading down from the largest scales ($k\lesssim8$) but the cascade is weaker than its initial magnitude.
The reduction in strength of the cascade on large scale is directly linked to the decay
of the large initial vortices.
Nevertheless, even at late times the overall energy balance is still dominated by the
largest scales, cf., the spectra shown in Fig.~\ref{fig:spectra_over_time} when taking into
account the $k^{4/3}$ compensation used in the plot.
Overall, while here the inertial range shrinks and becomes weaker (to a limited degree)
over time as the large scale modes lose energy, the dynamics within the inertial range 
is similar to driven turbulence simulations.

\subsection{Comparison to previous results}
\label{sec:comp-prev}

In general, our results in the weakly compressible MHD regime  are in agreement with the $\alpha \simeq -2$ spectrum reported by previous works on the TG vortex  in \cite{pouquetDynamicsUnforcedTurbulence2010,leeLackUniversalityDecaying2010,dallasOriginsEnsuremath2Spectrum2013-fixed,dallasStructuresDynamicsSmall2013} in the imcompressible MHD regime using the insulating magnetic field configuration.
We see the same $\alpha \simeq -2$ spectrum early in the evolution before $t=2T$, which corresponds to the time period near maximum energy dissipation that these other studies focused on.
In all cases that we simulated the spectra became shallower at later times, independent of the initial magnetization (whereas these other works focused on  $E_U/E_B=1$, i.e., $\mathcal{M}_{A,0}=1$, configurations, which are in
good agreement with the \MsMa{0.2}{1.0} simulation presented here, see top left panel of Fig.~\ref{fig:energy_histories}).
As noted by \cite{dallasOriginsEnsuremath2Spectrum2013-fixed}, the $\alpha \simeq -2$ spectrum is likely due to discontinuities in a small volume of the flow that can be disrupted by symmetry breaking at either large or small scales \cite{dallasSymmetryBreakingDecaying2013}.
According to \cite{dallasSymmetryBreakingDecaying2013}, a simulated Taylor-Green vortex with sufficiently high Reynolds number should show symmetry breaking at the small scales at late times in the evolution, causing a break from the $-2$ power law at large wavenumbers.
Since our simulations do not impose symmetries on the flow, this is a possible explanation for the observed behavior.
However, we see an $\alpha \simeq -4/3$ inertial range scaling at late times, instead of the $\alpha \simeq -2$ and $\alpha \simeq -5/3$ broken power law theorized by \cite{dallasSymmetryBreakingDecaying2013}.

Finally, work done in \cite{leeLackUniversalityDecaying2010,brachetIdealEvolutionMagnetohydrodynamic2013,dallasStructuresDynamicsSmall2013} shows that the behavior of the magnetic field and spectra changes with the initial magnetic field configurations.
With the insulating initial magnetic fields that we use, the vorticity begins parallel to the magnetic field.  This facilitates the early energy flux from kinetic to magnetic energy.
The insulating case tends towards stronger large magnetic fields compared to the other magnetic field configurations.
Both of the other initial magnetic fields result in different energy spectra, with the conducting magnetic field setup leading to a $k^{-3/2}$ spectra and the alternative insulating field setup leading to spectra interpreted as either a $k^{-5/3}$ or $k^{-2}$ spectra as argued by \cite{leeLackUniversalityDecaying2010} and \cite{dallasStructuresDynamicsSmall2013} respectively.

\subsection{Implication of results}
\label{sec:magnetic_fields_dominate_evolution}

In all of our simulations, we see magnetic fields and effects facilitated by the magnetic fields dominating the evolution of the decaying turbulence, even when the initial kinetic energy exceeds the magnetic energy by a factor of 100 in the \Ma{10} simulations.
Energy transfer from kinetic to magnetic energy via tension and energy transfer within the magnetic energy far exceed energy flux via the kinetic turbulent cascade at later times.
Energy transfer from kinetic to magnetic energy at earlier times leads to the magnetic energy dominating over kinetic energy in all cases in both total magnitude as well as in terms of the scale-wise budget, cf., magnetic versus kinetic energy spectra.
This is similar to what has been found in incompressible \citep{Alexakis2005} and compressible simulations \citep{grete_energy_2017,Grete2020slope} of driven turbulence.
Thus, even in intermittently-driven systems one can expect the magnetic field to significantly influence the dynamics after a few dynamical times.

Our simulations exhibit a magnetic energy spectra with a measurable power law after the turbulent flow is realized.
The inertial range is short, from approximately $k=10$ to $k=32$, due to the resolution of these simulations.
Nevertheless, within this region we can reasonably fit a power law to both the kinetic and magnetic spectra, which is often not possible in driven turbulence simulation without a dynamically relevant mean magnetic field, cf.,~Sec.~\ref{sec:magnetic_spectral_index}.
Thus, freely evolving and driven turbulence simulations complement each other and both are required to disentangle environmental from intrinsic effects.

From an observational point of view, we demonstrated that the spectral indices evolve over time and fluctuate even for similar parameters.
Therefore, the derived spectral indices from observation (e.g., velocity maps in astrophysics), which represent individual snapshots in time, need to be interpreted with care when trying to infer the ``nature'' of turbulence (e.g., Kolmogorov or Burgers) in the object of interest.

Finally, the observed nonlocal energy transfer has implications on the dynamical development of small scale structures from intermittent or singular energy injection events.
Within the context of natural astrophysical and terrestrial plasmas, the nonlocal energy transfer from kinetic to magnetic energies suggests that small magnetic field structures develop before small scale kinetic structures.

\subsection{Limitations} 

While our analysis showed that the results are generally robust (e.g., with respect to varying the fitting range in the spectral indices or varying range in the definition of scale-local in the energy transfers), higher resolution simulations are desirable.
With higher resolution in an implicit large eddy simulation (ILES) the dynamic range is increased and, thus, the effective Reynolds numbers of the simulated plasma are raised.

Similarly, due to the nature of ILES the effective magnetic Prandtl number in all simulations
is $\Pm \simeq 1 $.
However, in natural systems (both astrophysical and terrestrial/experimental) $\Pm$ is either
$\gg 1$ or $\ll 1$, motivating the exploration of these regimes in the future as well.

All of our simulations started with subsonic initial conditions, leaving the supersonic regime unexplored.
The additional shocks, discontinuities, and strong density variations that may arise in a supersonic flow could alter the energy transfer as the flow transitions into turbulence.
In the simulations we present here, the Mach number generally did not significantly affect the growth and behavior of the turbulence.
In a supersonic flow, however, the transitory effects such as the nonlocal energy transfer and inverse cascade may be altered or suppressed in addition to generally richer dynamics related to compressive effects and effective space-filling of turbulent structures \citep{Federrath2013a}.

Figure 5 indicates that the spectral index of both the kinetic and magnetic energy cascades evolves as a function of magnetic field strength (i.e., initial $\mathcal{M}_A$.)
It is unclear whether there is a threshold of $\mathcal{M}_A$ above which the spectra become shallower, or whether there is a continuum of behavior as the initial $\mathcal{M}_A$ is increased.  
While we would like to engage in a more thorough exploration of the dependence of these behaviors on $\mathcal{M}_A$, the simulations in question are computationally expensive and it is infeasible to do so at present.  Exploration of this transition is a promising venue for future work.
Finally, the shell decomposition used here to study energy transfer has been shown to violate the inviscid criterion for decomposing scales in the compressible regime \citep{Zhao2018}.
However, this only pertains to flows with significant density variations and, thus, 
is effectively irrelevant for the subsonic simulations presented here.

\section{Conclusions}
\label{sec:conclusions}

We have presented in this work nine simulations of the Taylor-Green vortex using the insulating magnetic field setup from \cite{leeParadigmaticFlowSmallscale2008} to study magnetized decaying turbulence in the compressible ideal MHD regime using the finite volume code \kathena.
As a first for the Taylor-Green vortex, we have also presented an energy transfer analysis to show the movement of energy between scales and energy reservoirs as facilitated via different mechanisms.   
Our key results are as follows:

\begin{itemize}
    \item Magnetic fields significantly affect the evolution of the decaying turbulence, regardless of initial field strength. Energy flux from kinetic energy to magnetic energy leads to the magnetic energy dominating the energy budget, even in simulations where the magnetic energy is initially very small. 
   
    \item The Taylor-Green vortex simulations explored here display a
power law in both the kinetic and magnetic energy spectra with a
measurable spectral index, which is in contrast with the lack of a power law in the magnetic energy spectrum seen
in driven turbulence calculations without a significant mean field.
    \item Decaying turbulent flows do not exhibit a spectral index
that is  constant in time  in either the kinetic nor magnetic energy
reservoirs -- these spectra continually evolve over time. 
The spectral
indices of the kinetic and magnetic energies become comparable and roughly constant around $1-2$ dynamical times after the magnetic energy has become dominant.  This can happen as early as $t=2T$ when the initial magnetic energy equals initial the kinetic energy, and as late as $t=5T$ when initial kinetic energy exceeds the magnetic by a factor of $100$.
For simulations with more initial kinetic energy than magnetic energy, the spectral indices reach a rough constant slightly steeper than $\alpha \simeq -4/3$.
    \item Before the turbulent flow fully develops, an inverse cascade
within the kinetic and magnetic energy reservoirs is intermittently observed.  This intermittent behavior moves energy from smaller
scales to larger scales, and is possible when the magnetic energy is comparable to the kinetic energy.
 \item Analysis of energy transfer within and between reservoirs indicates that within fully-developed turbulence, the cross-scale flux of energy in both the kinetic and magnetic cascades are dominated by energy transfer mediated by the magnetic field.
    \item Magnetic tension facilitates nonlocal transfer from larger
scales in the kinetic energy to smaller scales in the magnetic energy,
and is particularly prominent in simulations where the magnetic
field is initially weak.
\end{itemize}

\section{Acknowledgements}
This research is part of the Blue Waters sustained-petascale computing project, which is supported by the National Science Foundation (awards OCI-0725070 and ACI-1238993) the State of Illinois, and as of December, 2019, the National Geospatial-Intelligence Agency. Blue Waters is a joint effort of the University of Illinois at Urbana-Champaign and its National Center for Supercomputing Applications.
Exploratory simulations used the Blue Waters Supercomputer \cite{Bode2013,Kramer2015}.
The authors acknowledge the Texas Advanced Computing Center (TACC) at The University of Texas at Austin for providing HPC resources that have contributed to the research results reported within this paper.
The simulations presented here used the Stampede 2 Supercomputer.
FWG acknowledges support from the 2019 Blue Waters Graduate Fellowship. 
BWO acknowledges support from NSF grants no. AST-1517908 and AST-1908109, and NASA ATP grant 80NSSC18K1105.
PG and BWO  acknowledge funding by NASA ATP grant NNX15AP39G.
This work used the Extreme Science and Engineering Discovery Environment (XSEDE), which is supported by National Science Foundation grant number ACI-1548562, through allocation AST190021.  It also used the resources of the Michigan State University High Performance Computing Center, which is operated by the MSU Institute for Cyber-Enabled Research. 
\kathena \cite{greteKAthenaPerformancePortable2021}, \Athenapp \cite{stoneAthenaAdaptiveMesh2020}, \kokkos \citep{carteredwardsKokkosEnablingManycore2014}, and
\texttt{yt} \citep{turkYtMulticodeAnalysis2011} are developed by a large number of independent researchers from
numerous institutions around the world. Their commitment to open science
has helped make this work possible.

\providecommand{\noopsort}[1]{}\providecommand{\singleletter}[1]{#1}%


\begin{thebibliography}{34}%
\makeatletter
\providecommand \@ifxundefined [1]{%
 \@ifx{#1\undefined}
}%
\providecommand \@ifnum [1]{%
 \ifnum #1\expandafter \@firstoftwo
 \else \expandafter \@secondoftwo
 \fi
}%
\providecommand \@ifx [1]{%
 \ifx #1\expandafter \@firstoftwo
 \else \expandafter \@secondoftwo
 \fi
}%
\providecommand \natexlab [1]{#1}%
\providecommand \enquote  [1]{``#1''}%
\providecommand \bibnamefont  [1]{#1}%
\providecommand \bibfnamefont [1]{#1}%
\providecommand \citenamefont [1]{#1}%
\providecommand \href@noop [0]{\@secondoftwo}%
\providecommand \href [0]{\begingroup \@sanitize@url \@href}%
\providecommand \@href[1]{\@@startlink{#1}\@@href}%
\providecommand \@@href[1]{\endgroup#1\@@endlink}%
\providecommand \@sanitize@url [0]{\catcode `\\12\catcode `\$12\catcode
  `\&12\catcode `\#12\catcode `\^12\catcode `\_12\catcode `\%12\relax}%
\providecommand \@@startlink[1]{}%
\providecommand \@@endlink[0]{}%
\providecommand \url  [0]{\begingroup\@sanitize@url \@url }%
\providecommand \@url [1]{\endgroup\@href {#1}{\urlprefix }}%
\providecommand \urlprefix  [0]{URL }%
\providecommand \Eprint [0]{\href }%
\providecommand \doibase [0]{https://doi.org/}%
\providecommand \selectlanguage [0]{\@gobble}%
\providecommand \bibinfo  [0]{\@secondoftwo}%
\providecommand \bibfield  [0]{\@secondoftwo}%
\providecommand \translation [1]{[#1]}%
\providecommand \BibitemOpen [0]{}%
\providecommand \bibitemStop [0]{}%
\providecommand \bibitemNoStop [0]{.\EOS\space}%
\providecommand \EOS [0]{\spacefactor3000\relax}%
\providecommand \BibitemShut  [1]{\csname bibitem#1\endcsname}%
\let\auto@bib@innerbib\@empty
\bibitem [{\citenamefont {Norman}\ and\ \citenamefont
  {Bryan}(1999)}]{normanClusterTurbulence1999}%
  \BibitemOpen
  \bibfield  {author} {\bibinfo {author} {\bibfnamefont {M.~L.}\ \bibnamefont
  {Norman}}\ and\ \bibinfo {author} {\bibfnamefont {G.~L.}\ \bibnamefont
  {Bryan}},\ }\bibfield  {title} {\bibinfo {title} {Cluster turbulence},\ }in\
  \href {https://doi.org/10.1007/BFb0106425} {\emph {\bibinfo {booktitle} {The
  {{Radio Galaxy Messier}} 87}}},\ \bibinfo {series and number} {Lecture
  {{Notes}} in {{Physics}}},\ \bibinfo {editor} {edited by\ \bibinfo {editor}
  {\bibfnamefont {H.-J.}\ \bibnamefont {R{\"o}ser}}\ and\ \bibinfo {editor}
  {\bibfnamefont {K.}~\bibnamefont {Meisenheimer}}}\ (\bibinfo  {publisher}
  {{Springer}},\ \bibinfo {address} {{Berlin, Heidelberg}},\ \bibinfo {year}
  {1999})\ pp.\ \bibinfo {pages} {106--115}\BibitemShut {NoStop}%
\bibitem [{\citenamefont {Larson}(1981)}]{larsonTurbulenceStarFormation1981}%
  \BibitemOpen
  \bibfield  {author} {\bibinfo {author} {\bibfnamefont {R.~B.}\ \bibnamefont
  {Larson}},\ }\bibfield  {title} {\bibinfo {title} {Turbulence and star
  formation in molecular clouds},\ }\href
  {https://doi.org/10.1093/mnras/194.4.809} {\bibfield  {journal} {\bibinfo
  {journal} {Monthly Notices of the Royal Astronomical Society}\ }\textbf
  {\bibinfo {volume} {194}},\ \bibinfo {pages} {809} (\bibinfo {year}
  {1981})}\BibitemShut {NoStop}%
\bibitem [{\citenamefont {Britzen}\ \emph {et~al.}(2017)\citenamefont
  {Britzen}, \citenamefont {Fendt}, \citenamefont {Eckart},\ and\ \citenamefont
  {Karas}}]{britzenNewView872017}%
  \BibitemOpen
  \bibfield  {author} {\bibinfo {author} {\bibfnamefont {S.}~\bibnamefont
  {Britzen}}, \bibinfo {author} {\bibfnamefont {C.}~\bibnamefont {Fendt}},
  \bibinfo {author} {\bibfnamefont {A.}~\bibnamefont {Eckart}},\ and\ \bibinfo
  {author} {\bibfnamefont {V.}~\bibnamefont {Karas}},\ }\bibfield  {title}
  {\bibinfo {title} {A new view on the {{M}} 87 jet origin: {{Turbulent}}
  loading leading to large-scale episodic wiggling},\ }\href
  {https://doi.org/10.1051/0004-6361/201629469} {\bibfield  {journal} {\bibinfo
   {journal} {Astronomy \& Astrophysics}\ }\textbf {\bibinfo {volume} {601}},\
  \bibinfo {pages} {A52} (\bibinfo {year} {2017})}\BibitemShut {NoStop}%
\bibitem [{\citenamefont {Korpi}\ \emph {et~al.}(1999)\citenamefont {Korpi},
  \citenamefont {Brandenburg}, \citenamefont {Shukurov}, \citenamefont
  {Tuominen},\ and\ \citenamefont
  {Nordlund}}]{korpiSupernovaregulatedInterstellarMedium1999-fixed}%
  \BibitemOpen
  \bibfield  {author} {\bibinfo {author} {\bibfnamefont {M.~J.}\ \bibnamefont
  {Korpi}}, \bibinfo {author} {\bibfnamefont {A.}~\bibnamefont {Brandenburg}},
  \bibinfo {author} {\bibfnamefont {A.}~\bibnamefont {Shukurov}}, \bibinfo
  {author} {\bibfnamefont {I.}~\bibnamefont {Tuominen}},\ and\ \bibinfo
  {author} {\bibfnamefont {A.}~\bibnamefont {Nordlund}},\ }\bibfield  {title}
  {\bibinfo {title} {A {{Supernova}}-regulated {{Interstellar Medium}}:
  {{Simulations}} of the {{Turbulent Multiphase Medium}}},\ }\href
  {https://doi.org/10.1086/311954} {\bibfield  {journal} {\bibinfo  {journal}
  {The Astrophysical Journal}\ }\textbf {\bibinfo {volume} {514}},\ \bibinfo
  {pages} {L99} (\bibinfo {year} {1999})}\BibitemShut {NoStop}%
\bibitem [{\citenamefont {Rudakov}\ and\ \citenamefont
  {Sudan}(1997)}]{rudakovMHDTurbulenceRadiating1997a}%
  \BibitemOpen
  \bibfield  {author} {\bibinfo {author} {\bibfnamefont {L.~I.}\ \bibnamefont
  {Rudakov}}\ and\ \bibinfo {author} {\bibfnamefont {R.~N.}\ \bibnamefont
  {Sudan}},\ }\bibfield  {title} {\bibinfo {title} {{{MHD}} turbulence in
  radiating intense {{Z}}-pinches},\ }\href
  {https://doi.org/10.1016/S0370-1573(96)00062-2} {\bibfield  {journal}
  {\bibinfo  {journal} {Physics Reports}\ }\bibinfo {series} {Turbulence and
  {{Intermittency}} in {{Plasmas}}},\ \textbf {\bibinfo {volume} {283}},\
  \bibinfo {pages} {253} (\bibinfo {year} {1997})}\BibitemShut {NoStop}%
\bibitem [{\citenamefont {Kroupp}\ \emph {et~al.}(2018)\citenamefont {Kroupp},
  \citenamefont {Stambulchik}, \citenamefont {Starobinets}, \citenamefont
  {Osin}, \citenamefont {Fisher}, \citenamefont {Alumot}, \citenamefont
  {Maron}, \citenamefont {Davidovits}, \citenamefont {Fisch},\ and\
  \citenamefont {Fruchtman}}]{krouppTurbulentStagnationPinch2018-fixed}%
  \BibitemOpen
  \bibfield  {author} {\bibinfo {author} {\bibfnamefont {E.}~\bibnamefont
  {Kroupp}}, \bibinfo {author} {\bibfnamefont {E.}~\bibnamefont {Stambulchik}},
  \bibinfo {author} {\bibfnamefont {A.}~\bibnamefont {Starobinets}}, \bibinfo
  {author} {\bibfnamefont {D.}~\bibnamefont {Osin}}, \bibinfo {author}
  {\bibfnamefont {V.~I.}\ \bibnamefont {Fisher}}, \bibinfo {author}
  {\bibfnamefont {D.}~\bibnamefont {Alumot}}, \bibinfo {author} {\bibfnamefont
  {Y.}~\bibnamefont {Maron}}, \bibinfo {author} {\bibfnamefont
  {S.}~\bibnamefont {Davidovits}}, \bibinfo {author} {\bibfnamefont {N.~J.}\
  \bibnamefont {Fisch}},\ and\ \bibinfo {author} {\bibfnamefont
  {A.}~\bibnamefont {Fruchtman}},\ }\bibfield  {title} {\bibinfo {title}
  {Turbulent stagnation in a $z$-pinch plasma},\ }\href
  {https://doi.org/10.1103/PhysRevE.97.013202} {\bibfield  {journal} {\bibinfo
  {journal} {Physical Review E}\ }\textbf {\bibinfo {volume} {97}},\ \bibinfo
  {pages} {013202} (\bibinfo {year} {2018})}\BibitemShut {NoStop}%
\bibitem [{\citenamefont {Grete}\ \emph {et~al.}(2018)\citenamefont {Grete},
  \citenamefont {O'Shea},\ and\ \citenamefont
  {Beckwith}}]{greteMatterForceSystematic2018}%
  \BibitemOpen
  \bibfield  {author} {\bibinfo {author} {\bibfnamefont {P.}~\bibnamefont
  {Grete}}, \bibinfo {author} {\bibfnamefont {B.~W.}\ \bibnamefont {O'Shea}},\
  and\ \bibinfo {author} {\bibfnamefont {K.}~\bibnamefont {Beckwith}},\
  }\bibfield  {title} {\bibinfo {title} {As a {{Matter}} of
  {{Force}}\textemdash{{Systematic Biases}} in {{Idealized Turbulence
  Simulations}}},\ }\href {https://doi.org/10.3847/2041-8213/aac0f5} {\bibfield
   {journal} {\bibinfo  {journal} {The Astrophysical Journal}\ }\textbf
  {\bibinfo {volume} {858}},\ \bibinfo {pages} {L19} (\bibinfo {year}
  {2018})}\BibitemShut {NoStop}%
\bibitem [{\citenamefont {Taylor}\ and\ \citenamefont
  {Green}(1937)}]{taylorMechanismProductionSmall1937}%
  \BibitemOpen
  \bibfield  {author} {\bibinfo {author} {\bibfnamefont {G.~I.}\ \bibnamefont
  {Taylor}}\ and\ \bibinfo {author} {\bibfnamefont {A.~E.}\ \bibnamefont
  {Green}},\ }\bibfield  {title} {\bibinfo {title} {Mechanism of the
  {{Production}} of {{Small Eddies}} from {{Large Ones}}},\ }\href
  {https://doi.org/10.1098/rspa.1937.0036} {\bibfield  {journal} {\bibinfo
  {journal} {Proceedings of the Royal Society of London Series A}\ }\textbf
  {\bibinfo {volume} {158}},\ \bibinfo {pages} {499} (\bibinfo {year}
  {1937})}\BibitemShut {NoStop}%
\bibitem [{\citenamefont {Wang}\ \emph {et~al.}(2013)\citenamefont {Wang},
  \citenamefont {Fidkowski}, \citenamefont {Abgrall}, \citenamefont {Bassi},
  \citenamefont {Caraeni}, \citenamefont {Cary}, \citenamefont {Deconinck},
  \citenamefont {Hartmann}, \citenamefont {Hillewaert}, \citenamefont {Huynh}
  \emph {et~al.}}]{wang2013high}%
  \BibitemOpen
  \bibfield  {author} {\bibinfo {author} {\bibfnamefont {Z.~J.}\ \bibnamefont
  {Wang}}, \bibinfo {author} {\bibfnamefont {K.}~\bibnamefont {Fidkowski}},
  \bibinfo {author} {\bibfnamefont {R.}~\bibnamefont {Abgrall}}, \bibinfo
  {author} {\bibfnamefont {F.}~\bibnamefont {Bassi}}, \bibinfo {author}
  {\bibfnamefont {D.}~\bibnamefont {Caraeni}}, \bibinfo {author} {\bibfnamefont
  {A.}~\bibnamefont {Cary}}, \bibinfo {author} {\bibfnamefont {H.}~\bibnamefont
  {Deconinck}}, \bibinfo {author} {\bibfnamefont {R.}~\bibnamefont {Hartmann}},
  \bibinfo {author} {\bibfnamefont {K.}~\bibnamefont {Hillewaert}}, \bibinfo
  {author} {\bibfnamefont {H.~T.}\ \bibnamefont {Huynh}}, \emph {et~al.},\
  }\bibfield  {title} {\bibinfo {title} {High-order cfd methods: current status
  and perspective},\ }\href
  {https://onlinelibrary.wiley.com/doi/abs/10.1002/fld.3767} {\bibfield
  {journal} {\bibinfo  {journal} {International Journal for Numerical Methods
  in Fluids}\ }\textbf {\bibinfo {volume} {72}},\ \bibinfo {pages} {811}
  (\bibinfo {year} {2013})}\BibitemShut {NoStop}%
\bibitem [{\citenamefont {Brachet}\ \emph {et~al.}(1983)\citenamefont
  {Brachet}, \citenamefont {Meiron}, \citenamefont {Orszag}, \citenamefont
  {Nickel}, \citenamefont {Morf},\ and\ \citenamefont
  {Frisch}}]{brachetSmallscaleStructureTaylor1983}%
  \BibitemOpen
  \bibfield  {author} {\bibinfo {author} {\bibfnamefont {M.~E.}\ \bibnamefont
  {Brachet}}, \bibinfo {author} {\bibfnamefont {D.~I.}\ \bibnamefont {Meiron}},
  \bibinfo {author} {\bibfnamefont {S.~A.}\ \bibnamefont {Orszag}}, \bibinfo
  {author} {\bibfnamefont {B.~G.}\ \bibnamefont {Nickel}}, \bibinfo {author}
  {\bibfnamefont {R.~H.}\ \bibnamefont {Morf}},\ and\ \bibinfo {author}
  {\bibfnamefont {U.}~\bibnamefont {Frisch}},\ }\bibfield  {title} {\bibinfo
  {title} {Small-scale structure of the {{Taylor}}\textendash{{Green}}
  vortex},\ }\href {https://doi.org/10.1017/S0022112083001159} {\bibfield
  {journal} {\bibinfo  {journal} {Journal of Fluid Mechanics}\ }\textbf
  {\bibinfo {volume} {130}},\ \bibinfo {pages} {411} (\bibinfo {year}
  {1983})}\BibitemShut {NoStop}%
\bibitem [{\citenamefont {Lee}\ \emph {et~al.}(2008)\citenamefont {Lee},
  \citenamefont {Brachet}, \citenamefont {Pouquet}, \citenamefont {Mininni},\
  and\ \citenamefont {Rosenberg}}]{leeParadigmaticFlowSmallscale2008}%
  \BibitemOpen
  \bibfield  {author} {\bibinfo {author} {\bibfnamefont {E.}~\bibnamefont
  {Lee}}, \bibinfo {author} {\bibfnamefont {M.~E.}\ \bibnamefont {Brachet}},
  \bibinfo {author} {\bibfnamefont {A.}~\bibnamefont {Pouquet}}, \bibinfo
  {author} {\bibfnamefont {P.~D.}\ \bibnamefont {Mininni}},\ and\ \bibinfo
  {author} {\bibfnamefont {D.}~\bibnamefont {Rosenberg}},\ }\bibfield  {title}
  {\bibinfo {title} {Paradigmatic flow for small-scale magnetohydrodynamics:
  {{Properties}} of the ideal case and the collision of current sheets},\
  }\href {https://doi.org/10.1103/PhysRevE.78.066401} {\bibfield  {journal}
  {\bibinfo  {journal} {Physical Review E}\ }\textbf {\bibinfo {volume} {78}},\
  \bibinfo {pages} {066401} (\bibinfo {year} {2008})}\BibitemShut {NoStop}%
\bibitem [{\citenamefont {Lee}\ \emph {et~al.}(2010)\citenamefont {Lee},
  \citenamefont {Brachet}, \citenamefont {Pouquet}, \citenamefont {Mininni},\
  and\ \citenamefont {Rosenberg}}]{leeLackUniversalityDecaying2010}%
  \BibitemOpen
  \bibfield  {author} {\bibinfo {author} {\bibfnamefont {E.}~\bibnamefont
  {Lee}}, \bibinfo {author} {\bibfnamefont {M.~E.}\ \bibnamefont {Brachet}},
  \bibinfo {author} {\bibfnamefont {A.}~\bibnamefont {Pouquet}}, \bibinfo
  {author} {\bibfnamefont {P.~D.}\ \bibnamefont {Mininni}},\ and\ \bibinfo
  {author} {\bibfnamefont {D.}~\bibnamefont {Rosenberg}},\ }\bibfield  {title}
  {\bibinfo {title} {On the lack of universality in decaying
  magnetohydrodynamic turbulence},\ }\href
  {https://doi.org/10.1103/PhysRevE.81.016318} {\bibfield  {journal} {\bibinfo
  {journal} {Physical Review E}\ }\textbf {\bibinfo {volume} {81}},\ \bibinfo
  {pages} {016318} (\bibinfo {year} {2010})},\ \Eprint
  {https://arxiv.org/abs/0906.2506} {arXiv:0906.2506} \BibitemShut {NoStop}%
\bibitem [{\citenamefont {Pouquet}\ \emph {et~al.}(2010)\citenamefont
  {Pouquet}, \citenamefont {Lee}, \citenamefont {Brachet}, \citenamefont
  {Mininni},\ and\ \citenamefont
  {Rosenberg}}]{pouquetDynamicsUnforcedTurbulence2010}%
  \BibitemOpen
  \bibfield  {author} {\bibinfo {author} {\bibfnamefont {A.}~\bibnamefont
  {Pouquet}}, \bibinfo {author} {\bibfnamefont {E.}~\bibnamefont {Lee}},
  \bibinfo {author} {\bibfnamefont {M.~E.}\ \bibnamefont {Brachet}}, \bibinfo
  {author} {\bibfnamefont {P.~D.}\ \bibnamefont {Mininni}},\ and\ \bibinfo
  {author} {\bibfnamefont {D.}~\bibnamefont {Rosenberg}},\ }\bibfield  {title}
  {\bibinfo {title} {The dynamics of unforced turbulence at high {{Reynolds}}
  number for {{Taylor}}-{{Green}} vortices generalized to {{MHD}}},\ }\href
  {https://doi.org/10.1080/03091920903304080} {\bibfield  {journal} {\bibinfo
  {journal} {Geophysical and Astrophysical Fluid Dynamics}\ }\textbf {\bibinfo
  {volume} {104}},\ \bibinfo {pages} {115} (\bibinfo {year}
  {2010})}\BibitemShut {NoStop}%
\bibitem [{\citenamefont {Brachet}\ \emph {et~al.}(2013)\citenamefont
  {Brachet}, \citenamefont {Bustamante}, \citenamefont {Krstulovic},
  \citenamefont {Mininni}, \citenamefont {Pouquet},\ and\ \citenamefont
  {Rosenberg}}]{brachetIdealEvolutionMagnetohydrodynamic2013}%
  \BibitemOpen
  \bibfield  {author} {\bibinfo {author} {\bibfnamefont {M.~E.}\ \bibnamefont
  {Brachet}}, \bibinfo {author} {\bibfnamefont {M.~D.}\ \bibnamefont
  {Bustamante}}, \bibinfo {author} {\bibfnamefont {G.}~\bibnamefont
  {Krstulovic}}, \bibinfo {author} {\bibfnamefont {P.~D.}\ \bibnamefont
  {Mininni}}, \bibinfo {author} {\bibfnamefont {A.}~\bibnamefont {Pouquet}},\
  and\ \bibinfo {author} {\bibfnamefont {D.}~\bibnamefont {Rosenberg}},\
  }\bibfield  {title} {\bibinfo {title} {Ideal evolution of magnetohydrodynamic
  turbulence when imposing {{Taylor}}-{{Green}} symmetries},\ }\href
  {https://doi.org/10.1103/PhysRevE.87.013110} {\bibfield  {journal} {\bibinfo
  {journal} {Physical Review E}\ }\textbf {\bibinfo {volume} {87}},\ \bibinfo
  {pages} {013110} (\bibinfo {year} {2013})}\BibitemShut {NoStop}%
\bibitem [{\citenamefont {Dallas}\ and\ \citenamefont
  {Alexakis}(2013{\natexlab{a}})}]{dallasOriginsEnsuremath2Spectrum2013-fixed}%
  \BibitemOpen
  \bibfield  {author} {\bibinfo {author} {\bibfnamefont {V.}~\bibnamefont
  {Dallas}}\ and\ \bibinfo {author} {\bibfnamefont {A.}~\bibnamefont
  {Alexakis}},\ }\bibfield  {title} {\bibinfo {title} {Origins of the
  ${k}^{-2}$ spectrum in decaying {{Taylor}}-{{Green}} magnetohydrodynamic
  turbulent flows},\ }\href {https://doi.org/10.1103/PhysRevE.88.053014}
  {\bibfield  {journal} {\bibinfo  {journal} {Physical Review E}\ }\textbf
  {\bibinfo {volume} {88}},\ \bibinfo {pages} {053014} (\bibinfo {year}
  {2013}{\natexlab{a}})}\BibitemShut {NoStop}%
\bibitem [{\citenamefont {Dallas}\ and\ \citenamefont
  {Alexakis}(2013{\natexlab{b}})}]{dallasStructuresDynamicsSmall2013}%
  \BibitemOpen
  \bibfield  {author} {\bibinfo {author} {\bibfnamefont {V.}~\bibnamefont
  {Dallas}}\ and\ \bibinfo {author} {\bibfnamefont {A.}~\bibnamefont
  {Alexakis}},\ }\bibfield  {title} {\bibinfo {title} {Structures and dynamics
  of small scales in decaying magnetohydrodynamic turbulence},\ }\href
  {https://doi.org/10.1063/1.4824195} {\bibfield  {journal} {\bibinfo
  {journal} {Physics of Fluids}\ }\textbf {\bibinfo {volume} {25}},\ \bibinfo
  {pages} {105106} (\bibinfo {year} {2013}{\natexlab{b}})},\ \Eprint
  {https://arxiv.org/abs/1304.0695} {arXiv:1304.0695} \BibitemShut {NoStop}%
\bibitem [{\citenamefont {Dallas}\ and\ \citenamefont
  {Alexakis}(2013{\natexlab{c}})}]{dallasSymmetryBreakingDecaying2013}%
  \BibitemOpen
  \bibfield  {author} {\bibinfo {author} {\bibfnamefont {V.}~\bibnamefont
  {Dallas}}\ and\ \bibinfo {author} {\bibfnamefont {A.}~\bibnamefont
  {Alexakis}},\ }\bibfield  {title} {\bibinfo {title} {Symmetry breaking of
  decaying magnetohydrodynamic {{Taylor}}-{{Green}} flows and consequences for
  universality},\ }\href {https://doi.org/10.1103/PhysRevE.88.063017}
  {\bibfield  {journal} {\bibinfo  {journal} {Physical Review E}\ }\textbf
  {\bibinfo {volume} {88}},\ \bibinfo {pages} {063017} (\bibinfo {year}
  {2013}{\natexlab{c}})}\BibitemShut {NoStop}%
\bibitem [{\citenamefont {Vahala}\ \emph {et~al.}(2008)\citenamefont {Vahala},
  \citenamefont {Keating}, \citenamefont {Soe}, \citenamefont {Yepez},
  \citenamefont {Vahala}, \citenamefont {Carter},\ and\ \citenamefont
  {Ziegeler}}]{vahalaMHDTurbulenceStudies2008-fixed}%
  \BibitemOpen
  \bibfield  {author} {\bibinfo {author} {\bibfnamefont {G.}~\bibnamefont
  {Vahala}}, \bibinfo {author} {\bibfnamefont {B.}~\bibnamefont {Keating}},
  \bibinfo {author} {\bibfnamefont {M.}~\bibnamefont {Soe}}, \bibinfo {author}
  {\bibfnamefont {J.}~\bibnamefont {Yepez}}, \bibinfo {author} {\bibfnamefont
  {L.}~\bibnamefont {Vahala}}, \bibinfo {author} {\bibfnamefont
  {J.}~\bibnamefont {Carter}},\ and\ \bibinfo {author} {\bibfnamefont
  {S.}~\bibnamefont {Ziegeler}},\ }\bibfield  {title} {\bibinfo {title} {{{MHD
  Turbulence Studies}} using {{Lattice Boltzmann Algorithms}}},\ }\href
  {https://global-sci.org/intro/article_detail/cicp/7808.html} {\bibfield
  {journal} {\bibinfo  {journal} {Commun. Comput. Phys.}\ ,\ \bibinfo {pages}
  {23}} (\bibinfo {year} {2008})}\BibitemShut {NoStop}%
\bibitem [{\citenamefont {Yang}\ \emph {et~al.}(2016)\citenamefont {Yang},
  \citenamefont {Shi}, \citenamefont {Wan}, \citenamefont {Matthaeus},\ and\
  \citenamefont {Chen}}]{Yang2016}%
  \BibitemOpen
  \bibfield  {author} {\bibinfo {author} {\bibfnamefont {Y.}~\bibnamefont
  {Yang}}, \bibinfo {author} {\bibfnamefont {Y.}~\bibnamefont {Shi}}, \bibinfo
  {author} {\bibfnamefont {M.}~\bibnamefont {Wan}}, \bibinfo {author}
  {\bibfnamefont {W.~H.}\ \bibnamefont {Matthaeus}},\ and\ \bibinfo {author}
  {\bibfnamefont {S.}~\bibnamefont {Chen}},\ }\bibfield  {title} {\bibinfo
  {title} {Energy cascade and its locality in compressible magnetohydrodynamic
  turbulence},\ }\href {https://doi.org/10.1103/PhysRevE.93.061102} {\bibfield
  {journal} {\bibinfo  {journal} {Phys. Rev. E}\ }\textbf {\bibinfo {volume}
  {93}},\ \bibinfo {pages} {061102} (\bibinfo {year} {2016})}\BibitemShut
  {NoStop}%
\bibitem [{\citenamefont {Grete}\ \emph {et~al.}(2017)\citenamefont {Grete},
  \citenamefont {O'Shea}, \citenamefont {Beckwith}, \citenamefont {Schmidt},\
  and\ \citenamefont {Christlieb}}]{grete_energy_2017}%
  \BibitemOpen
  \bibfield  {author} {\bibinfo {author} {\bibfnamefont {P.}~\bibnamefont
  {Grete}}, \bibinfo {author} {\bibfnamefont {B.~W.}\ \bibnamefont {O'Shea}},
  \bibinfo {author} {\bibfnamefont {K.}~\bibnamefont {Beckwith}}, \bibinfo
  {author} {\bibfnamefont {W.}~\bibnamefont {Schmidt}},\ and\ \bibinfo {author}
  {\bibfnamefont {A.}~\bibnamefont {Christlieb}},\ }\bibfield  {title}
  {\bibinfo {title} {Energy transfer in compressible magnetohydrodynamic
  turbulence},\ }\href {https://doi.org/10.1063/1.4990613} {\bibfield
  {journal} {\bibinfo  {journal} {Physics of Plasmas}\ }\textbf {\bibinfo
  {volume} {24}},\ \bibinfo {pages} {092311} (\bibinfo {year}
  {2017})}\BibitemShut {NoStop}%
\bibitem [{\citenamefont {Grete}\ \emph
  {et~al.}(2021{\natexlab{a}})\citenamefont {Grete}, \citenamefont {Glines},\
  and\ \citenamefont {O'Shea}}]{greteKAthenaPerformancePortable2021}%
  \BibitemOpen
  \bibfield  {author} {\bibinfo {author} {\bibfnamefont {P.}~\bibnamefont
  {Grete}}, \bibinfo {author} {\bibfnamefont {F.~W.}\ \bibnamefont {Glines}},\
  and\ \bibinfo {author} {\bibfnamefont {B.~W.}\ \bibnamefont {O'Shea}},\
  }\bibfield  {title} {\bibinfo {title} {K-{{Athena}}: {{A Performance Portable
  Structured Grid Finite Volume Magnetohydrodynamics Code}}},\ }\href
  {https://doi.org/10.1109/TPDS.2020.3010016} {\bibfield  {journal} {\bibinfo
  {journal} {IEEE Transactions on Parallel and Distributed Systems}\ }\textbf
  {\bibinfo {volume} {32}},\ \bibinfo {pages} {85} (\bibinfo {year}
  {2021}{\natexlab{a}})},\ \Eprint {https://arxiv.org/abs/1905.04341}
  {arXiv:1905.04341} \BibitemShut {NoStop}%
\bibitem [{\citenamefont {Stone}\ \emph {et~al.}(2020)\citenamefont {Stone},
  \citenamefont {Tomida}, \citenamefont {White},\ and\ \citenamefont
  {Felker}}]{stoneAthenaAdaptiveMesh2020}%
  \BibitemOpen
  \bibfield  {author} {\bibinfo {author} {\bibfnamefont {J.~M.}\ \bibnamefont
  {Stone}}, \bibinfo {author} {\bibfnamefont {K.}~\bibnamefont {Tomida}},
  \bibinfo {author} {\bibfnamefont {C.~J.}\ \bibnamefont {White}},\ and\
  \bibinfo {author} {\bibfnamefont {K.~G.}\ \bibnamefont {Felker}},\ }\bibfield
   {title} {\bibinfo {title} {The {{Athena}}++ {{Adaptive Mesh Refinement
  Framework}}: {{Design}} and {{Magnetohydrodynamic Solvers}}},\ }\href
  {https://doi.org/10.3847/1538-4365/ab929b} {\bibfield  {journal} {\bibinfo
  {journal} {The Astrophysical Journal Supplement Series}\ }\textbf {\bibinfo
  {volume} {249}},\ \bibinfo {pages} {4} (\bibinfo {year} {2020})}\BibitemShut
  {NoStop}%
\bibitem [{\citenamefont {Carter~Edwards}\ \emph {et~al.}(2014)\citenamefont
  {Carter~Edwards}, \citenamefont {Trott},\ and\ \citenamefont
  {Sunderland}}]{carteredwardsKokkosEnablingManycore2014}%
  \BibitemOpen
  \bibfield  {author} {\bibinfo {author} {\bibfnamefont {H.}~\bibnamefont
  {Carter~Edwards}}, \bibinfo {author} {\bibfnamefont {C.~R.}\ \bibnamefont
  {Trott}},\ and\ \bibinfo {author} {\bibfnamefont {D.}~\bibnamefont
  {Sunderland}},\ }\bibfield  {title} {\bibinfo {title} {Kokkos: {{Enabling}}
  manycore performance portability through polymorphic memory access
  patterns},\ }\href {https://doi.org/10.1016/j.jpdc.2014.07.003} {\bibfield
  {journal} {\bibinfo  {journal} {Journal of Parallel and Distributed
  Computing}\ }\bibinfo {series} {Domain-{{Specific Languages}} and
  {{High}}-{{Level Frameworks}} for {{High}}-{{Performance Computing}}},\
  \textbf {\bibinfo {volume} {74}},\ \bibinfo {pages} {3202} (\bibinfo {year}
  {2014})}\BibitemShut {NoStop}%
\bibitem [{\citenamefont {Stone}\ and\ \citenamefont
  {Gardiner}(2009)}]{Stone2009}%
  \BibitemOpen
  \bibfield  {author} {\bibinfo {author} {\bibfnamefont {J.~M.}\ \bibnamefont
  {Stone}}\ and\ \bibinfo {author} {\bibfnamefont {T.}~\bibnamefont
  {Gardiner}},\ }\bibfield  {title} {\bibinfo {title} {A simple unsplit godunov
  method for multidimensional mhd},\ }\href
  {https://doi.org/https://doi.org/10.1016/j.newast.2008.06.003} {\bibfield
  {journal} {\bibinfo  {journal} {New Astronomy}\ }\textbf {\bibinfo {volume}
  {14}},\ \bibinfo {pages} {139 } (\bibinfo {year} {2009})}\BibitemShut
  {NoStop}%
\bibitem [{Note1()}]{Note1}%
  \BibitemOpen
  \bibinfo {note} {Note that other works such as \protect \citet
  {wang2013high,pouquetDynamicsUnforcedTurbulence2010} use a nondimensional
  time, $t^* = L/u_0$, in contrast to the dynamical time used
  here.}\BibitemShut {Stop}%
\bibitem [{\citenamefont {Alexakis}\ \emph {et~al.}(2005)\citenamefont
  {Alexakis}, \citenamefont {Mininni},\ and\ \citenamefont
  {Pouquet}}]{Alexakis2005}%
  \BibitemOpen
  \bibfield  {author} {\bibinfo {author} {\bibfnamefont {A.}~\bibnamefont
  {Alexakis}}, \bibinfo {author} {\bibfnamefont {P.~D.}\ \bibnamefont
  {Mininni}},\ and\ \bibinfo {author} {\bibfnamefont {A.}~\bibnamefont
  {Pouquet}},\ }\bibfield  {title} {\bibinfo {title} {Shell-to-shell energy
  transfer in magnetohydrodynamics. i. steady state turbulence},\ }\href
  {https://doi.org/10.1103/PhysRevE.72.046301} {\bibfield  {journal} {\bibinfo
  {journal} {Phys. Rev. E}\ }\textbf {\bibinfo {volume} {72}},\ \bibinfo
  {pages} {046301} (\bibinfo {year} {2005})}\BibitemShut {NoStop}%
\bibitem [{\citenamefont {Kida}\ and\ \citenamefont
  {Orszag}(1990)}]{kidaEnergySpectralDynamics1990}%
  \BibitemOpen
  \bibfield  {author} {\bibinfo {author} {\bibfnamefont {S.}~\bibnamefont
  {Kida}}\ and\ \bibinfo {author} {\bibfnamefont {S.~A.}\ \bibnamefont
  {Orszag}},\ }\bibfield  {title} {\bibinfo {title} {Energy and spectral
  dynamics in forced compressible turbulence},\ }\href
  {https://doi.org/10.1007/BF01065580} {\bibfield  {journal} {\bibinfo
  {journal} {Journal of Scientific Computing}\ }\textbf {\bibinfo {volume}
  {5}},\ \bibinfo {pages} {85} (\bibinfo {year} {1990})}\BibitemShut {NoStop}%
\bibitem [{\citenamefont {Domaradzki}\ \emph {et~al.}(2010)\citenamefont
  {Domaradzki}, \citenamefont {Teaca},\ and\ \citenamefont
  {Carati}}]{Domaradzki2010}%
  \BibitemOpen
  \bibfield  {author} {\bibinfo {author} {\bibfnamefont {J.~A.}\ \bibnamefont
  {Domaradzki}}, \bibinfo {author} {\bibfnamefont {B.}~\bibnamefont {Teaca}},\
  and\ \bibinfo {author} {\bibfnamefont {D.}~\bibnamefont {Carati}},\
  }\bibfield  {title} {\bibinfo {title} {Locality properties of the energy flux
  in magnetohydrodynamic turbulence},\ }\href
  {https://doi.org/10.1063/1.3431227} {\bibfield  {journal} {\bibinfo
  {journal} {Physics of Fluids}\ }\textbf {\bibinfo {volume} {22}},\ \bibinfo
  {pages} {051702} (\bibinfo {year} {2010})},\ \Eprint
  {https://arxiv.org/abs/http://dx.doi.org/10.1063/1.3431227}
  {http://dx.doi.org/10.1063/1.3431227} \BibitemShut {NoStop}%
\bibitem [{\citenamefont {Grete}\ \emph
  {et~al.}(2021{\natexlab{b}})\citenamefont {Grete}, \citenamefont {O'Shea},\
  and\ \citenamefont {Beckwith}}]{Grete2020slope}%
  \BibitemOpen
  \bibfield  {author} {\bibinfo {author} {\bibfnamefont {P.}~\bibnamefont
  {Grete}}, \bibinfo {author} {\bibfnamefont {B.~W.}\ \bibnamefont {O'Shea}},\
  and\ \bibinfo {author} {\bibfnamefont {K.}~\bibnamefont {Beckwith}},\
  }\bibfield  {title} {\bibinfo {title} {As a matter of tension: Kinetic energy
  spectra in {MHD} turbulence},\ }\href
  {https://doi.org/10.3847/1538-4357/abdd22} {\bibfield  {journal} {\bibinfo
  {journal} {The Astrophysical Journal}\ }\textbf {\bibinfo {volume} {909}},\
  \bibinfo {pages} {148} (\bibinfo {year} {2021}{\natexlab{b}})}\BibitemShut
  {NoStop}%
\bibitem [{\citenamefont {Federrath}(2013)}]{Federrath2013a}%
  \BibitemOpen
  \bibfield  {author} {\bibinfo {author} {\bibfnamefont {C.}~\bibnamefont
  {Federrath}},\ }\bibfield  {title} {\bibinfo {title} {{On the universality of
  supersonic turbulence}},\ }\href {https://doi.org/10.1093/mnras/stt1644}
  {\bibfield  {journal} {\bibinfo  {journal} {Mon. Not. R. Astron. Soc.}\
  }\textbf {\bibinfo {volume} {436}},\ \bibinfo {pages} {1245} (\bibinfo {year}
  {2013})}\BibitemShut {NoStop}%
\bibitem [{\citenamefont {Zhao}\ and\ \citenamefont {Aluie}(2018)}]{Zhao2018}%
  \BibitemOpen
  \bibfield  {author} {\bibinfo {author} {\bibfnamefont {D.}~\bibnamefont
  {Zhao}}\ and\ \bibinfo {author} {\bibfnamefont {H.}~\bibnamefont {Aluie}},\
  }\bibfield  {title} {\bibinfo {title} {Inviscid criterion for decomposing
  scales},\ }\href {https://doi.org/10.1103/PhysRevFluids.3.054603} {\bibfield
  {journal} {\bibinfo  {journal} {Phys. Rev. Fluids}\ }\textbf {\bibinfo
  {volume} {3}},\ \bibinfo {pages} {054603} (\bibinfo {year}
  {2018})}\BibitemShut {NoStop}%
\bibitem [{\citenamefont {Bode}\ \emph {et~al.}(2013)\citenamefont {Bode},
  \citenamefont {Butler}, \citenamefont {Dunning}, \citenamefont {Hoefler},
  \citenamefont {Kramer}, \citenamefont {Gropp},\ and\ \citenamefont
  {Hwu}}]{Bode2013}%
  \BibitemOpen
  \bibfield  {author} {\bibinfo {author} {\bibfnamefont {B.}~\bibnamefont
  {Bode}}, \bibinfo {author} {\bibfnamefont {M.}~\bibnamefont {Butler}},
  \bibinfo {author} {\bibfnamefont {T.}~\bibnamefont {Dunning}}, \bibinfo
  {author} {\bibfnamefont {T.}~\bibnamefont {Hoefler}}, \bibinfo {author}
  {\bibfnamefont {W.}~\bibnamefont {Kramer}}, \bibinfo {author} {\bibfnamefont
  {W.}~\bibnamefont {Gropp}},\ and\ \bibinfo {author} {\bibfnamefont {W.-m.}\
  \bibnamefont {Hwu}},\ }\bibfield  {title} {\bibinfo {title} {The {B}lue
  {W}aters super-system for super-science},\ }in\ \href
  {https://www.taylorfrancis.com/books/e/9781466568358} {\emph {\bibinfo
  {booktitle} {Contemporary High Performance Computing}}},\ \bibinfo {series
  and number} {Chapman {\&} Hall/CRC Computational Science}\ (\bibinfo
  {publisher} {Chapman and Hall/CRC},\ \bibinfo {year} {2013})\ pp.\ \bibinfo
  {pages} {339--366}\BibitemShut {NoStop}%
\bibitem [{\citenamefont {Kramer}\ \emph {et~al.}(2015)\citenamefont {Kramer},
  \citenamefont {Butler}, \citenamefont {Bauer}, \citenamefont {Chadalavada},\
  and\ \citenamefont {Mendes}}]{Kramer2015}%
  \BibitemOpen
  \bibfield  {author} {\bibinfo {author} {\bibfnamefont {W.}~\bibnamefont
  {Kramer}}, \bibinfo {author} {\bibfnamefont {M.}~\bibnamefont {Butler}},
  \bibinfo {author} {\bibfnamefont {G.}~\bibnamefont {Bauer}}, \bibinfo
  {author} {\bibfnamefont {K.}~\bibnamefont {Chadalavada}},\ and\ \bibinfo
  {author} {\bibfnamefont {C.}~\bibnamefont {Mendes}},\ }\bibfield  {title}
  {\bibinfo {title} {{Blue Waters Parallel I/O Storage Sub-system}},\ }in\
  \href@noop {} {\emph {\bibinfo {booktitle} {High Performance Parallel
  I/O}}},\ \bibinfo {editor} {edited by\ \bibinfo {editor} {\bibnamefont
  {Prabhat}}\ and\ \bibinfo {editor} {\bibfnamefont {Q.}~\bibnamefont
  {Koziol}}}\ (\bibinfo  {publisher} {CRC Publications, Taylor and Francis
  Group},\ \bibinfo {year} {2015})\ pp.\ \bibinfo {pages} {17--32}\BibitemShut
  {NoStop}%
\bibitem [{\citenamefont {Turk}\ \emph {et~al.}(2011)\citenamefont {Turk},
  \citenamefont {Smith}, \citenamefont {Oishi}, \citenamefont {Skory},
  \citenamefont {Skillman}, \citenamefont {Abel},\ and\ \citenamefont
  {Norman}}]{turkYtMulticodeAnalysis2011}%
  \BibitemOpen
  \bibfield  {author} {\bibinfo {author} {\bibfnamefont {M.~J.}\ \bibnamefont
  {Turk}}, \bibinfo {author} {\bibfnamefont {B.~D.}\ \bibnamefont {Smith}},
  \bibinfo {author} {\bibfnamefont {J.~S.}\ \bibnamefont {Oishi}}, \bibinfo
  {author} {\bibfnamefont {S.}~\bibnamefont {Skory}}, \bibinfo {author}
  {\bibfnamefont {S.~W.}\ \bibnamefont {Skillman}}, \bibinfo {author}
  {\bibfnamefont {T.}~\bibnamefont {Abel}},\ and\ \bibinfo {author}
  {\bibfnamefont {M.~L.}\ \bibnamefont {Norman}},\ }\bibfield  {title}
  {\bibinfo {title} {Yt: {{A Multi}}-code {{Analysis Toolkit}} for
  {{Astrophysical Simulation Data}}},\ }\href
  {https://doi.org/10.1088/0067-0049/192/1/9} {\bibfield  {journal} {\bibinfo
  {journal} {The Astrophysical Journal Supplement Series}\ }\textbf {\bibinfo
  {volume} {192}},\ \bibinfo {pages} {9} (\bibinfo {year} {2011})}\BibitemShut
  {NoStop}%
\end{thebibliography}
\end{document}